# Systematic conformation-to-phenotype mapping via limited deep-sequencing of proteins

**Eugene Serebryany[1] *, Victor Y. Zhao[1], Kibum Park[1], Amir Bitran[1], Sunia A. Trauger[2], Bogdan Budnik[2], and Eugene I. Shakhnovich[1] ***
*[1]Department of Chemistry and Chemical Biology and [2]Center for Mass Spectrometry, Harvard University, Cambridge, MA*

## Summary

Non-native conformations drive protein misfolding diseases, complicate bioengineering efforts, and fuel molecular evolution. No current experimental technique is well-suited for elucidating them and their phenotypic effects. Especially intractable are the transient conformations populated by intrinsically disordered proteins. We describe an approach to systematically discover, stabilize, and purify native and non-native conformations, generated *in vitro* or *in vivo*, and directly link conformations to molecular, organismal, or evolutionary phenotypes. This approach involves high-throughput disulfide scanning (HTDS) of the entire protein. To reveal which disulfides trap which chromatographically resolvable conformers, we devised a deep-sequencing method for double-Cys variant libraries of proteins that precisely and simultaneously locates both Cys residues within each polypeptide. HTDS of the abundant *E. coli* periplasmic chaperone HdeA revealed distinct classes of disordered hydrophobic conformers with variable cytotoxicity depending on where the backbone was cross-linked. HTDS can bridge conformational and phenotypic landscapes for many proteins that function in disulfide-permissive environments.

## Introduction

Protein energy landscapes tend to be rugged, which means a protein molecule can adopt many distinct conformations. What effects do these diverse conformations have *in vivo*, and what biophysical properties account for those effects? Intrinsically disordered, fold-switching, and kinetically trapped proteins all populate multiple conformers physiologically[1-7] with distinct functions or phenotypes.[8-10] *In-vivo* conformational ensembles can also differ from *in-vitro* ones.[11-13] Misfolded conformers can be toxic *in vivo*,[14,15] but the biophysical mechanisms of such toxicity remain elusive, and experimental investigation of non-native, transient, and disordered conformations is very challenging. The conformation-phenotype mapping problem is especially acute for intrinsically disordered proteins (IDPs), which may fold up *in vivo*[16,17] or adopt a plethora of transient backbone conformations (**Figure 1a**).

Disulfide bonds encode 3D conformational constraints directly in a protein's sequence. Disulfides easily form in many environments *in vivo* and, thanks to the action of disulfide isomerases, generally do not perturb protein conformations but stabilize them.[18] Disulfide scanning mutagenesis (determining which double-Cys variants of a protein can form intramolecular disulfides) is a well-established technique for testing structural models – e.g., elucidating the angle and register between two α-helices.[19-22] However, the number of double-Cys variants scales as the square of polypeptide length, so scanning an entire protein has never been practical. We report a high-throughput disulfide scanning (**HTDS**) methodology that surmounts this key limitation, providing a qualitatively new capability: mapping a protein's conformational landscape onto phenotypic landscapes.

As proof of concept, we applied HTDS to a highly abundant *E. coli* periplasmic chaperone, HdeA. One of the few known IDPs in bacteria,[23,24] HdeA is a holdase that inhibits aggregation of periplasmic proteins as the bacterium passes through stomach acid on its way to the gut of a new host.[25-27] Its functional (low-pH) conformation is largely disordered,[26,28] but at neutral pH it has a folded core and a disordered N-terminal region and is dimeric.[25,29] A conserved disulfide links the only two Cys residues (18 and 66) in the native sequence; its reduction disorders the structure and inhibits chaperone activity *in vitro*,[28,30] though not entirely.[31] *In-vivo* consequences of a lost or shifted disulfide have not been investigated; we report that dose-dependent toxicity typically results.

We used DNA deep sequencing to determine allele toxicities of 1,453 double-Cys variants of HdeA on the C18S/C66S ("noC") background, expressed with the native periplasm-targeting sequence. We then devised a protein-level deep sequencing method to measure the *ex vivo* redox state and hydrophobicity of hundreds of the variant proteins and gained further biophysical insight for select variants via *in vitro* biophysical characterization and *in silico* modeling. Surprisingly, very few double-Cys variants rescued the allele toxicity of noC, but many greatly enhanced it. Allele toxicity was associated with increased protein hydrophobicity, evidence of protein misfolding stress *in vivo* (DnaK overexpression) and, ultimately, cell lysis. The expression strain of *E. coli* was Δ*hdea*, so the observed toxicity phenotype of HdeA mutants is evidence of a misfolding-induced gain of toxic function. We conclude that the folding of WT HdeA is dependent on its 18-66 disulfide, while a well-defined subset of ectopic disulfides favors disordered yet highly toxic backbone conformations. Our methodology can be directly applied to many other IDPs and paves the way for applications to larger and more folded proteins.

## Design

Crystallography, NMR, and cryoEM have revealed high-resolution native structures of ~50,000 proteins. However, mapping conformational to phenotypic landscapes *in vivo* is extremely challenging. Proteome-wide limited proteolysis-mass spectrometry (LP/MS) experiments suggest many proteins are disordered *in vitro* but not *ex vivo*,[16] but the method lacks resolution beyond whether a protein or domain is folded. In-cell NMR can yield insights into protein conformation,[32,33] but it entails significant perturbations to the cells, such as electroporation, and needs to be coupled to a method like HTDS to be able to study transient or rare non-native conformers. More importantly, purely structural methods cannot map distinct conformations to their phenotypic effects. Conversely, deep mutational scanning can reveal complex *in-vivo* mutational phenotypic landscapes in sequence space,[14,34-36] but it does not directly probe protein conformation. Intermediate, transient, and disordered conformations also do not leave clear signatures in the multiple sequence alignments on which computational methods like AlphaFold and RosettaFold depend. Other methods, including our MCPU, can map conformational landscapes computationally,[37-42] but the dearth of experimental data means they must be trained on native conformations only, and validation of the computational predictions has to rest on very limited experimental findings. Our aim (**Figure 1b**) is for HTDS to become a versatile method capable of mapping both native and non-native conformations, *in vitro* and *in vivo*, in monomeric and complexed states, and furthermore able to stabilize distinct conformers for downstream phenotypic or biophysical characterization.

To our knowledge, this study is the first implementation of HTDS even at the DNA level (via amplicon deep sequencing of pooled double-Cys mutant libraries). However, only single-molecule, long-read deep sequencing of *proteins* can reveal which double-Cys variants in a pooled library formed disulfides and which did not, or which ones are enriched in which chromatographic fraction. Despite impressive recent progress,[43] general single-molecule protein sequencing methods cannot yet rise to the HTDS challenge. We therefore devised our own protein deep-sequencing method that fulfills the limited but unique requirements of HTDS: precisely determining unknown positions of both Cys residues simultaneously within each polypeptide molecule in a pooled double-Cys variant library. We achieved this by site-specific polypeptide backbone cleavage at Cys positions via cyanylation-aminolysis chemistry,[44,45] splitting each double-Cys polypeptide into exactly three fragments; the unique termini of the middle fragment are assignable by MS/MS and correspond exactly to the two Cys positions in the intact protein (**Figure 1c**). This method minimizes mass spectral complexity and crowding, with no ambiguity over intra- vs. intermolecular disulfides, while adding a positive charge to the C-termini of cleavage fragments for greater ionizability and unambiguous identification.

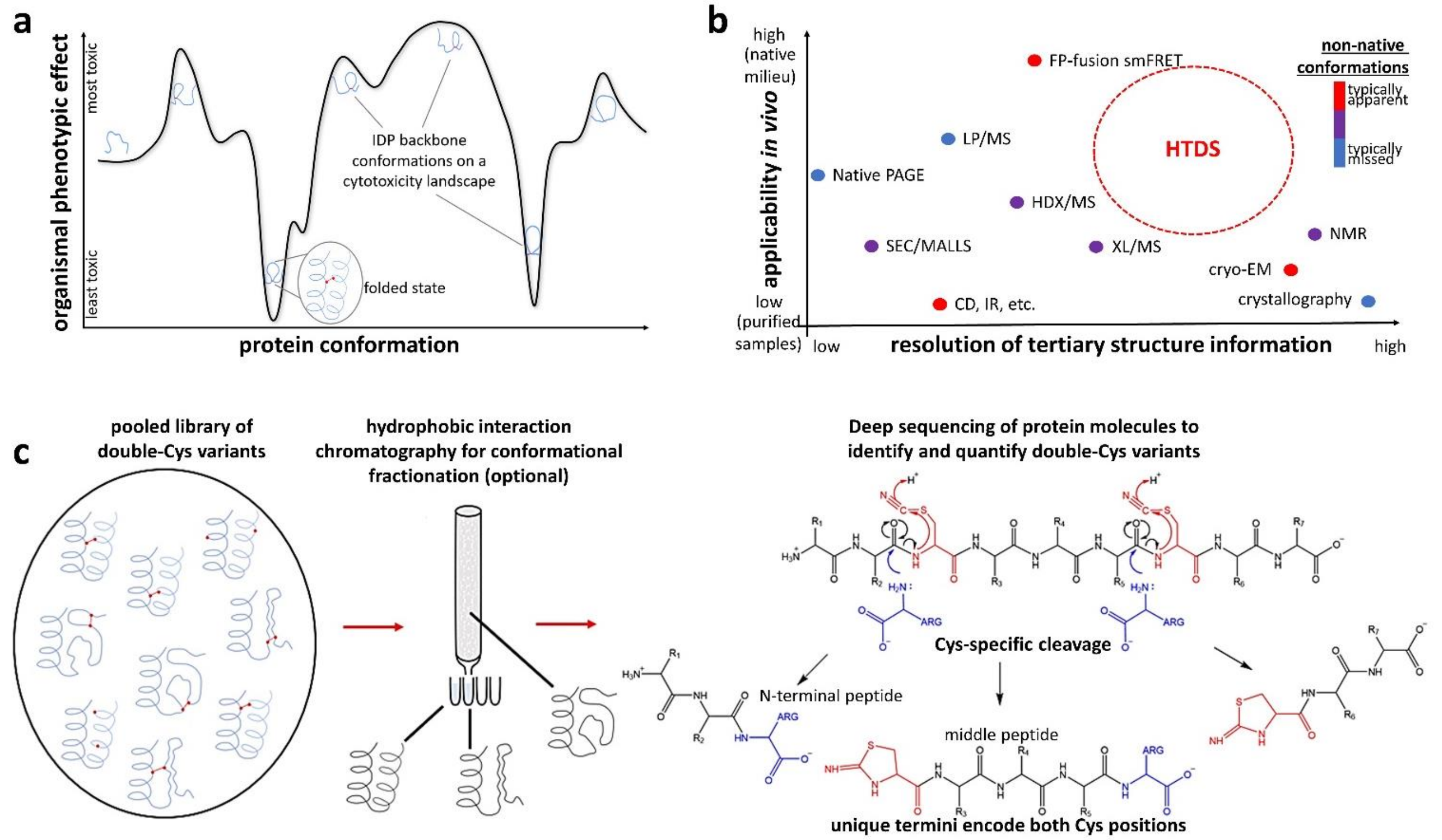

**Figure 1: Rationale and design for the HTDS method. (a)** In an IDP, distinct backbone conformations may produce distinct molecular and organismal phenotypes, such as cytotoxicity. **(b)** The niche for HTDS as a structural method: mapping both native and non-native conformers, both *in vitro* and *in vivo*, at moderate-to-high resolution. It is perhaps unique among all methods in its ability to connect specific conformational restraints to specific *in vivo* phenotypic effects. **(c)** The key enabling technology for HTDS is single-molecule deep sequencing of double-Cys protein libraries to locate both Cys in each polypeptide at once, thereby identifying and quantifying each variant across library fractions. We accomplish this via cyanylation-aminolysis: at pH 9, the amine of free Arg attacks the carbonyl immediately upstream of cyano-Cys – a good leaving group thanks to internal cyclization to iminothiazole (ITZ) – and is incorporated into the upstream fragment. The library may be fractionated by biophysical properties (e.g., hydrophobicity, as shown) or by chemical properties, such as redox status (i.e., disulfide-forming variants from non-disulfide-forming). Both approaches are demonstrated in this study.

HTDS has many advantages for probing conformational landscapes *in vivo* compared to conventional techniques. While limited proteolysis mass spectrometry (LP/MS) can only reveal whether a domain is folded or not (since everything unfolds after the protease cut), and hydrogen-deuterium exchange (HDX/MS) can only reveal whether a peptide is exposed or buried, HTDS can reveal tertiary contacts within a folded domain. Unlike crosslinking mass spectrometry (XL/MS), HTDS requires neither a chemical crosslinker, nor any unnatural amino acids, and data analysis is much simpler since crosslinks are encoded in linear peptides.

The information obtainable by HTDS depends on the properties of the protein. If the protein is completely rigid, then only disulfides consistent with the rigid native structure will form under physiological conditions, and that structure can then be calculated from the conformational restraints (illustrated in **Figure SI 1**); if it is completely soft (an IDP), then any disulfide can form, trapping that protein in all possible backbone conformations; and if the protein is part-rigid, part-soft, then the crosslinks that form *in vivo* will reflect the native conformation plus the set of easily-accessible non-native ones, and chromatographic fractionation will reveal which sets of disulfides are consistent with which. Disulfide-trapped non-native conformers are intriguing drug targets, e.g., for a novel class of antibiotics, while disulfide-crosslinked functional conformations can be useful for stabilizing biologic drugs and vaccines. Since they form *in vivo* under native conditions, disulfides can trap distinct protein conformations *in vivo*, making it possible to systematically study their phenotypic and even evolutionary effects.

Two caveats are worth mentioning upfront. First, expression levels of the double-Cys proteins can vary. This variation requires an *epistatic* definition of the measured *in vivo* phenotype, i.e., the effect of the *interaction* between two Cys substitutions corrected for the effects of the substitutions themselves. Second, in a natively dimeric protein like HdeA, there is a possibility of intermolecular disulfides. As discussed below, intermolecular disulfides were very rare in our study; however, they may form in studies of other proteins. Such cases can be resolved by size-exclusion chromatography (SEC) under chemically denaturing, but non-reducing, conditions. In such SEC runs, proteins that are covalently crosslinked can be collected as a separate elution peak and thereafter analyzed separately, if desired, to map subsets of sequence positions that enable intermolecular crosslinks.

## Results

### Ectopic disulfides in HdeA often result in cell lysis

In the following text, we name HdeA variants "XX/YY" to indicate sequence positions of their Cys residues. We generated a multi-Cys scanning library (mostly 2-Cys and 1-Cys variants) in the rhamnose-titratable pCK302 plasmid[46] and used the non-rhamnose-metabolizing BW25113 Δ*hdea E. coli*. We first measured growth curves of 11 variants (7 double-Cys variants picked randomly, plus WT, noC, and variants 32/66 and 40/65 chosen from preliminary studies), plus superfolder GFP (sGFP) as a control, in sealed 96-well plates with varying [rhamnose]. All HdeA variants had the native periplasmic targeting sequence; sGFP was expressed in the cytoplasm as a control for the cost of protein overexpression. Pronounced, dose-dependent dips appeared in the growth curves of most HdeA variants, but not WT, in early stationary phase (**Figure 2a**). All cultures in **Figure 2a** were inoculated identically and measured in parallel. Using tenfold smaller inocula altered the shapes of the growth curves, but variant-dependent toxicity remained, and the rank order of variants was similar (**Figure SI 2**).

All 11 HdeA variants were well expressed at high induction when inoculated from pH 8 LB starter cultures (**Figure 2b**). Expression from unbuffered LB starters was more variable (**Figure SI 3**) but still rhamnose-tunable (**Figure 2c**). We attribute the faster migration of WT (18/66), 11/73, 33/86, and 18/86 relative to noC, 32/66, 40/65, etc., in non-reducing SDS-PAGE (**Figure 2b**) to compaction of the denatured state by the longer-range disulfides. Expression level of 32/66 was below WT at low [rhamnose] but above WT at high [rhamnose] (**Figure 2d**), suggesting somewhat higher degradation propensity. Comparing total expression cultures to identical volumes of clarified supernatants of the same cultures yielded two unexpected observations (**Figure 2b**). First, HdeA was found mostly in the supernatant (unlike sGFP). We are not aware of prior studies of whether HdeA is exported from the cell. All constructs contained the periplasmic targeting sequence of native *E. coli* HdeA. Only the HdeA band was visible in WT HdeA supernatant, so the protein's export was not via cell lysis. Such "leakage" of overexpressed proteins has been observed before, though the mechanism is poorly understood.[47] By contrast, the many protein bands in supernatants of HdeA mutants indicated extensive cell lysis, which explains the dips in their growth curves (**Figure 2a**). Second, the lysogenic variants triggered strong overexpression of an endogenous 70 kD protein (marked on the gel in **Figure 2b**). LC/MS/MS assigned it as DnaK (7x more peptide spectrum matches (PSM) than the closest *E. coli* contaminant). Quantifying the band intensities (**Figure 2e**) suggested a DnaK/HdeA ratio converging to ~0.25 for both noC and 32/66. We conclude that cytotoxic HdeA variants triggered a protein-misfolding response in their host cells.

The die-off in stationary phase was typically followed by recovery (**Figure 2a**), which further investigation (**Figures SI 4 and SI 5**) revealed was driven by cells that had turned off the toxic variant's expression. Sanger sequencing of miniprepped plasmids from the recovered cultures ruled out plasmid loss and confirmed the HdeA coding sequences. More detailed study of the expression shut-off mechanism is beyond the scope of this work; here, we focused on the initial die-off as a convenient readout of variant allele toxicity.

### A landscape of allele toxicity for HdeA double-Cys variants

All seven randomly chosen double-Cys variants in **Figure 2b** caused some degree of cell lysis. We found it possible to directly measure variant lysogenicity in pooled cultures by PCR-amplifying HdeA coding sequences from plasmids released into the culture medium as a result of cell lysis and those retained in the cell pellet after centrifugation. This simple "killing assay" (**Figure 3a**) is conceptually similar to screening variants of a lytic gene of a phage: variants that efficiently lyse their host bacterium escape, and the others are retained. Thus, we can define a variant's "raw" allele toxicity as the ratio *R* of released to retained plasmid, determined as the proportion of paired-end Illumina reads for this variant in culture supernatant- vs. cell pellet-derived PCR amplicons.

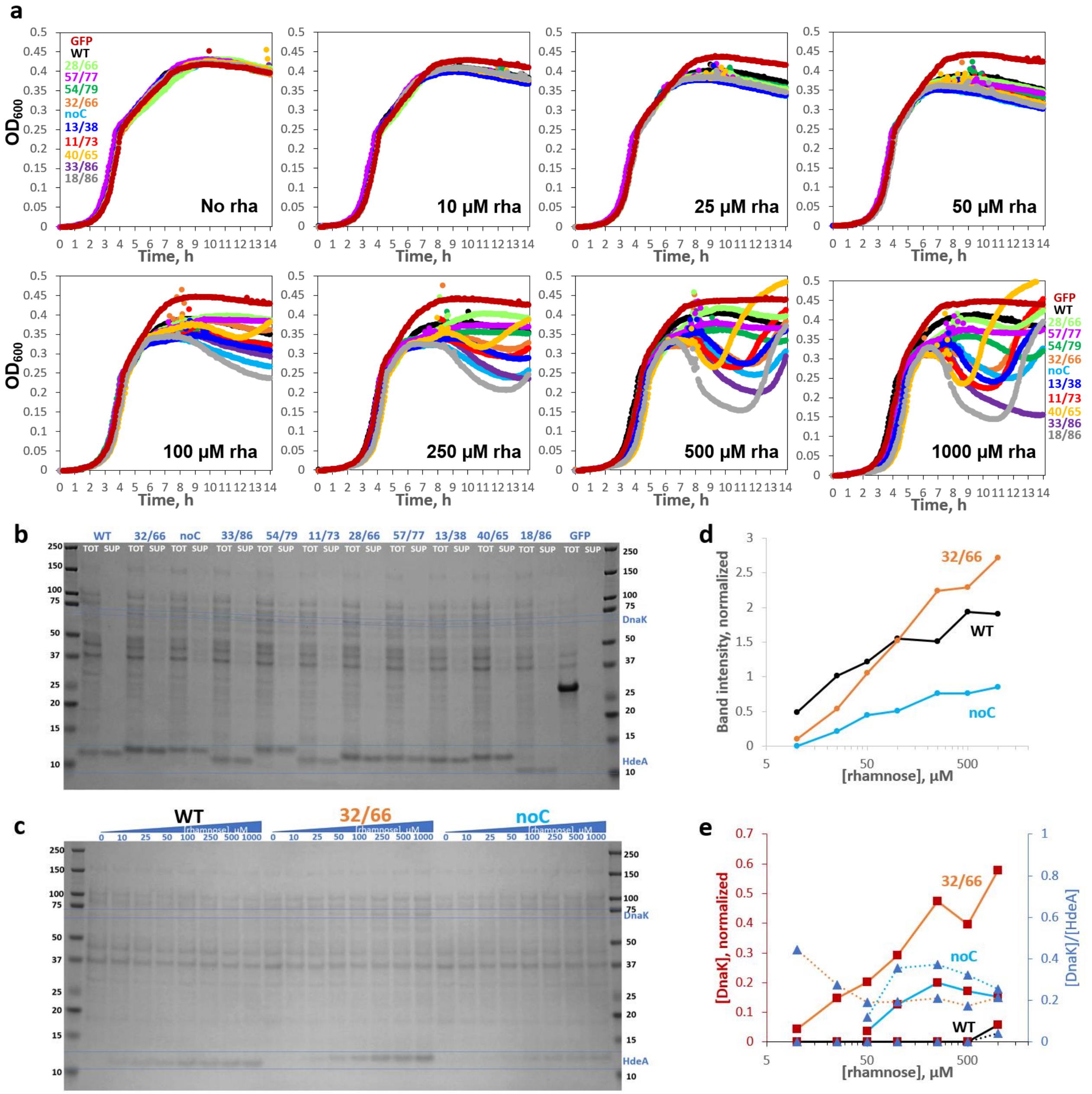

**Figure 2: HdeA variants exhibit varying levels of toxicity characterized by DnaK overexpression and cell lysis**. (**a**) Growth curves of HdeA variants in standard LB broth with varying [inducer], measured in parallel in the same 96-well plate, with all wells for each variant started from the same inoculum, showed variant- and [inducer]-dependent drops in $OD_{600}$. WT HdeA and sGFP did not. (**b**) Non-reducing SDS-PAGE of end-point samples from 2500 μM rhamnose cultures showed clear expression of all constructs, albeit weaker than sGFP. Variants with more sequence-distal Cys pairs migrated lower on the gel, consistent with greater compaction of the unfolded state by longer-range disulfides (see **Figure SI 6**); note rightward skew of the gel. Juxtaposing identical volumes of centrifuged cell culture supernatant ("SUP") and total culture ("TOT") samples showed 86±11% (mean±S.D.) of HdeA in SUP across variants. All mutants showed many cytoplasmic protein bands in SUP, indicating cell lysis, and DnaK overexpression (confirmed by MS/MS). DnaK's prominence in SUP indicated that it did not prevent lysis. (**c**) Non-reducing SDS-PAGE of the WT, 32/66, and noC constructs (TOT) as a function of [rhamnose]. (**d**) Quantitation of the HdeA bands in **c**, internally normalized to the abundant 37 kD band, showed a crossover between WT and 32/66 abundance. (**e**) Quantitation of the DnaK bands in panel **c** for 32/66 (*orange line*) and noC (*light-blue line*); observed [DnaK]/[HdeA] ratios converged to ~0.25.

For example, variant noC, the most abundant in our library, comprised 3.15 ± 0.01% (mean ± S.E.M.) of "retained" reads and 2.80 ± 0.02% of "released" reads, yielding *R* = 0.89. In total, 3,423 variants had at least 50 full paired-end Illumina reads on average across samples of four replicate cultures at 9.0 h post-induction (see Methods), including 1,453 double-Cys variants that passed the quality check for outliers. 76 of 85 theoretically possible single-Cys variants were likewise detected. All single-Cys variants combined comprised 24.73 ± 0.02% of "retained" and 23.76 ± 0.10% of "released" reads. All double-Cys variants combined comprised 46.03 ± 0.07% and 47.72 ± 0.17%, respectively. The balance was mostly sequences with three Cys or with a non-Cys mutation.

**Figure SI 7** shows the "raw" allele toxicity landscape (*R* values), but this metric requires two corrections. First, to reveal the effect of putative disulfides, we must isolate the effect of the *interaction* between the two Cys substitutions from effects of the individual mutations. We do this by defining *epistatic* allele toxicity *E* as follows:

$$E = \frac{R_{doubleCys}}{R_{Cys\#1} R_{Cys\#2}}$$

The landscape of this "epistatic" allele toxicity is also shown in **Figure SI 7**, plotted as a logarithm per convention;[48,49] it reveals that most of the apparently non-toxic variants on the "raw" toxicity landscape are attributable to the single-mutation effects, while the effects of their Cys-Cys interactions tend to be close to neutral. Second, because variation in absolute allele abundance in our library spans at least three orders of magnitude, the errors in abundance measurements also vary. To focus on the strongest and highest-confidence phenotypic effects, we therefore measure epistatic toxicity in units of #S.E.M. away from noC (for which *R* = *E* = 0.89). This is the landscape shown in **Figure 3b**.

Gratifyingly, either epistatic measure puts WT among the top 1% of least-toxic variants. Yet, the observed landscape differed markedly from WT's native contact map (**Figure 3c**). The cumulative average of *E* did show a clear signature of reduced toxicity for variants whose Cys residues had short native $C_\alpha$-$C_\alpha$ distances (**Figure 3d**), but the effect was surprisingly small, the cumulative averages barely rising above noC. Of the 44 constructs with native $C_\alpha$-$C_\alpha$ distances < 4 Å longer than for WT's disulfide-bonded residues 18 and 66, the least-toxic was WT itself (18/66, at +5.5 SEM relative to noC), followed by 17/66 (+2.4 SEM), and no other variant breached +2.0 SEM. Even variants 18/65 (-1.5 SEM) and 18/67 (-1.0 SEM) were more toxic than noC. Nor did shifting the disulfide by one helical turn rescue toxicity: 18/70 was at -0.5 SEM and 22/66 at -3.4. Meanwhile, swathes of double-Cys sequence space showed high epistatic allele toxicity, most notably a rough rectangle at positions 27-52 (Cys#1) X 63-66 (Cys#2). Thus, the native state of HdeA appears to be highly brittle: moving the disulfide from its native location – even when the new location remains plausible given the WT 3D structure – often causes toxicity to the organism.

Broadly speaking, the phenotypic effects of double-Cys mutations may be explained as the product of two factors: the intrinsic toxicity of the variant protein molecule and the variant's abundance, i.e., expression level. To verify each factor's contribution to overall toxicity, we used a random number generator to select 18 and 20 double-Cys variants from the least-toxic and most-toxic 10%, respectively, of the allele toxicity distribution (**Figure 3b**) and measured these variants' growth curves and expression levels as in **Figure 2**. Though the "non-toxic" variants had statistically lower expression levels than then the "toxic" ones (p = 0.0061 by the one-tailed Mann-Whitney U test), the difference was remarkably modest: 31 of the 38 randomly chosen variants had expression levels within a factor of 2 of the WT (**Figure 3e**). As shown in **Figure SI 8a-c**, expression levels showed a fair correlation with raw toxicity ($R^2$ = 0.40) but a much weaker correlation with mean-based epistatic ($R^2$ = 0.25) or SEM-based epistatic ($R^2$ = 0.16) toxicity as defined in **Figure 3b**. Allele toxicity correlated poorly with the magnitude of the observed dip in variant growth curves (**Figure SI 8d-f**). Growth curve shape measured by $OD_{600}$ is therefore not a good quantitative measure of allele toxicity; it is likely confounded by changes in cell size or shape (perhaps filamentation). The correlation between HdeA and DnaK expression should be considered significant given that estimates of DnaK levels are much noisier than those of HdeA levels (**Figure SI 8g-i**). We conclude that the epistatic definition of allele toxicity largely filters out variation in toxicity due to expression levels and foregrounds each variant's intrinsic toxicity.

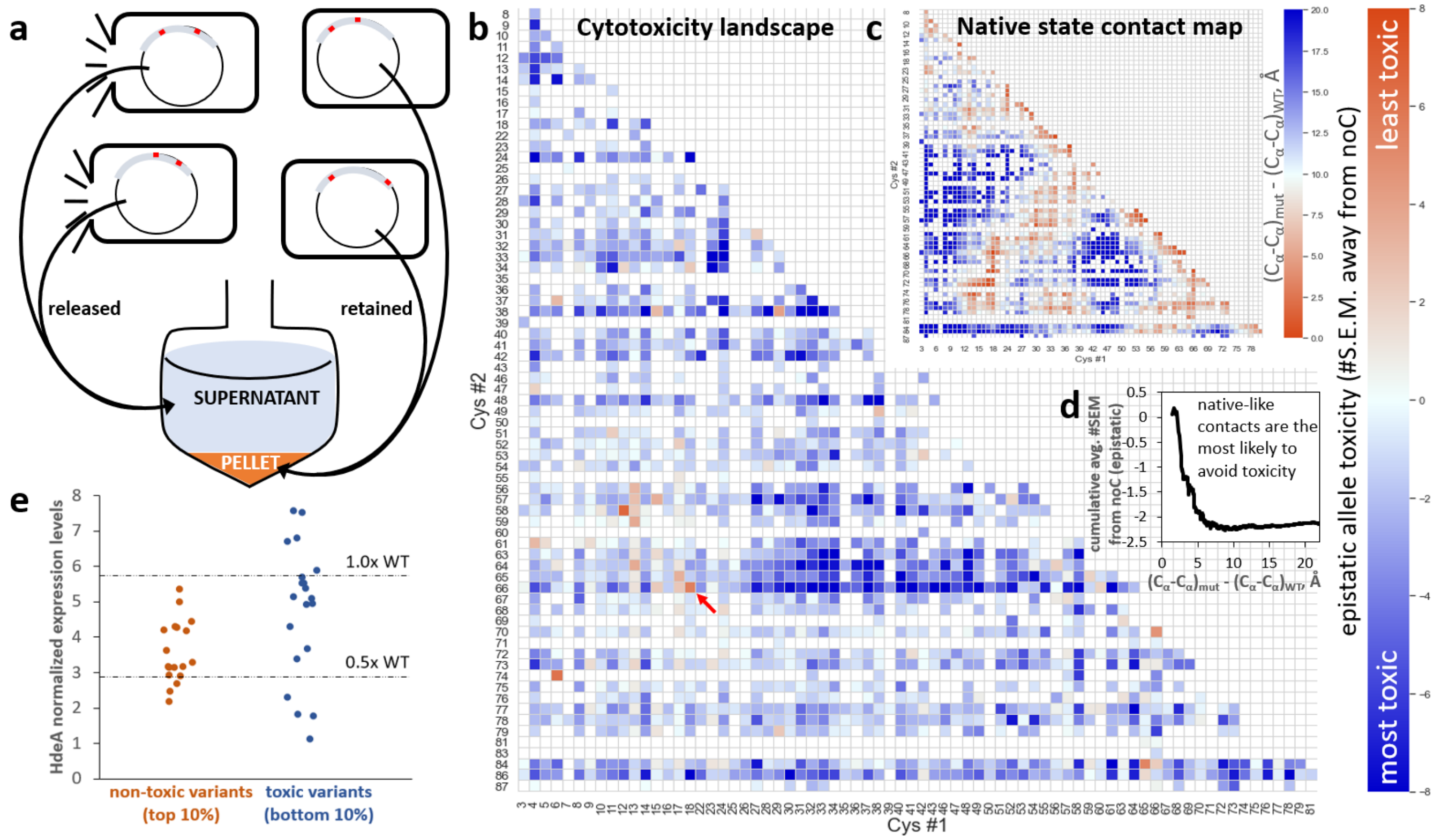

**Figure 3: Allele toxicity landscape of 1,453 double-Cys variants**. **(a)** A cartoon illustrating the experiment: HdeA variants were expressed in pooled culture; plasmids encoding lysogenic variants escaped into the culture medium as their host cells lysed, while those encoding less-toxic variants remained inside their host cells. During subsequent centrifugation, the former partitioned to the supernatant fraction, and the latter partitioned to the pellet. **(b)** An allele toxicity landscape for all 2-Cys variants that could be quantified by DNA deep sequencing in quadruplicate cultures grown with 1 mM rhamnose. Allele toxicity is defined as the difference of means of *E* between the given variant and noC, divided by the S.E.M. of the variant. Positive difference (*orange*) indicates less cell lysis than noC; negative difference (*blue*) indicates more lysis than noC. Very few variants were observed to be less toxic than noC. The WT gene (18/66, indicated by *red arrow*) ranked #3 out of 1,453 by this metric. **(c)** A pairwise contact map of $C_\alpha$-$C_\alpha$ distances relative to WT's, derived by averaging such distances in chain A and chain B of PDB ID 5WYO. **(d)** Cumulative average epistatic toxicity (expressed as in panel **b**) as a function of the pairwise native-state $C_\alpha$-$C_\alpha$ distance (*black*) from WT for all double-Cys variants with Cys at least 12 peptide bonds apart showed lower cytotoxicity for smaller $C_\alpha$-$C_\alpha$ distances. **(e)** Low-throughput measurements of expression levels of 38 randomly chosen double-Cys variants from the top and bottom 10% of the distribution of epistatic allele toxicities, defined as in panel **(b)**.

Although expression level differences in general had only a weak correlation with epistatic toxicity, they could explain some outlier cases. Thus, variants 12/58 and 6/74 were the only ones with even lower apparent allele toxicity than WT, and the former also had a clear island of non-toxicity around it in sequence space (**Figure 3b**). Even their growth curves differed from noC or the more toxic variants and resembled WT (**Figure SI 9**). The Cys pairs in 12/58 and 6/74 are too far apart for native-state intramolecular crosslinks, even considering N-terminal flexibility, and non-reducing SDS-PAGE showed no evidence of extensive intersubunit crosslinks, either (**Figure 2a, Figure SI 10**). Isotopically resolved intact-protein mass spectrometry confirmed intramolecular disulfide bonding (-2 Da. shift) in 12/58, along with WT and 32/66 (**Figure SI 11**); 6/74 could not be reliably measured due to extensive protein degradation during storage. Indeed, protein degradation was the most likely explanation for its lack of toxicity. At moderate induction, expression levels, DnaK overexpression, and cell lysis were all quite low for 6/74 and 12/58 (**Figure SI 10**). At very high induction, however (2.5 mM rhamnose, exceeding the 1 mM in **Figure 3b**), expression, DnaK, and cell lysis all returned. It appears, then, that some disulfides stabilize uniquely degradation-prone conformations, making those alleles appear non-toxic unless protein synthesis rates are increased to overmatch the increased degradation rates.

## Library-wide determination of variant redox status *ex vivo*

Strong epistasis suggests but does not prove disulfide bonding; weak epistasis may result from disulfides that stabilize benign conformations or simply fail to form. Proper interpretation of the DNA-level experiments requires knowing whether a given pair of Cys forms a disulfide *in vivo*. Therefore, we extracted the pooled protein library from the periplasm of the expressing cells, partially purified it (see Methods), and determined disulfide bonding status experimentally for all variants with peptides detectable at <10% false discovery rate (FDR). The library was split into treatment and reference samples: in the former, free thiols were blocked by N-ethyl maleimide (NEM), before reduction and cyanylation (as in **Figure 1b**); in the latter, NEM was not used, so every Cys was cyanylated. We confirmed by SDS-PAGE that NEM blocking of Cys residues prevents backbone cleavage at those sites (**Figure SI 12**).

While 1,453 double-Cys variants were detected at the DNA level (**Figure 3b**), only 1,154 were theoretically detectable at the protein level by virtue of having Cys residues 11-62 peptide bonds apart (see Limitations). Of those, 446 variants (~39% of theoretically possible) were detected at the protein level across all experiments in the present study: 284 in the periplasmic extract and 339 extracellularly, with an overlap of 177. Since HdeA is natively periplasmic, we used the periplasmic library to determine disulfide formation propensity *in vivo*. However, signals for 118 of the periplasmic variants were deemed too low to determine disulfide formation propensity (see **Figure SI 13**); the rest are shown in **Figure 4a**. Lack of protein-level detections should not be ascribed primarily to expression levels. Rather, observed protein-level abundance is the product of at least four factors: (1) allele abundance at the DNA level, (2) the expression level, (3) the degree of periplasmic localization, and (4) ionizability of MS/MS fragments. Allele abundance at the DNA level varied by three orders of magnitude because the library was generated via PCR-based mutagenesis (see Methods), and most protein variants that did not reach the abundance threshold for **Figure 4a** were encoded by low-abundance alleles (**Figure 4d**). The impact of DNA-level abundance is further illustrated by variants containing Cys66; those alleles were purposely enriched at the DNA level (see Methods), yielding 73% of theoretically possible protein-level coverage. Variation in protein expression levels was small compared to DNA-level variation (**Figure 3e**). Nearly half of protein variants detected extracellularly were not detected in the periplasmic extracts (see above); this could have been due to differences in periplasmic localization, but also to the extra fractionation step (see Methods). Finally, differences in ionizability among peptides are an intrinsic limitation of mass spectrometry.

Many variants in **Figure 4a** had strong disulfide formation propensities, suggesting a highly disordered or flexible underlying conformational ensemble. Disulfide bonding was not always all-or-none, consistent with the gel in **Figure 2b**, where the highly toxic 18/86 variant had a prominent downshifted band (expected given its long-range disulfide, see **Figure SI 6**) but also a faint band at the "noC" position. Intermolecular disulfides were rare in our experiments: the low-throughput study of expression levels of 38 randomly chosen variants (see Methods) showed dimer bands in non-reducing SDS-PAGE for only two variants, and even those were much fainter than the monomer bands. The average estimated *in vivo* disulfide bonding propensity across all variants was just below that of WT HdeA, which is fully disulfide-bonded (**Figure 4b,c**).

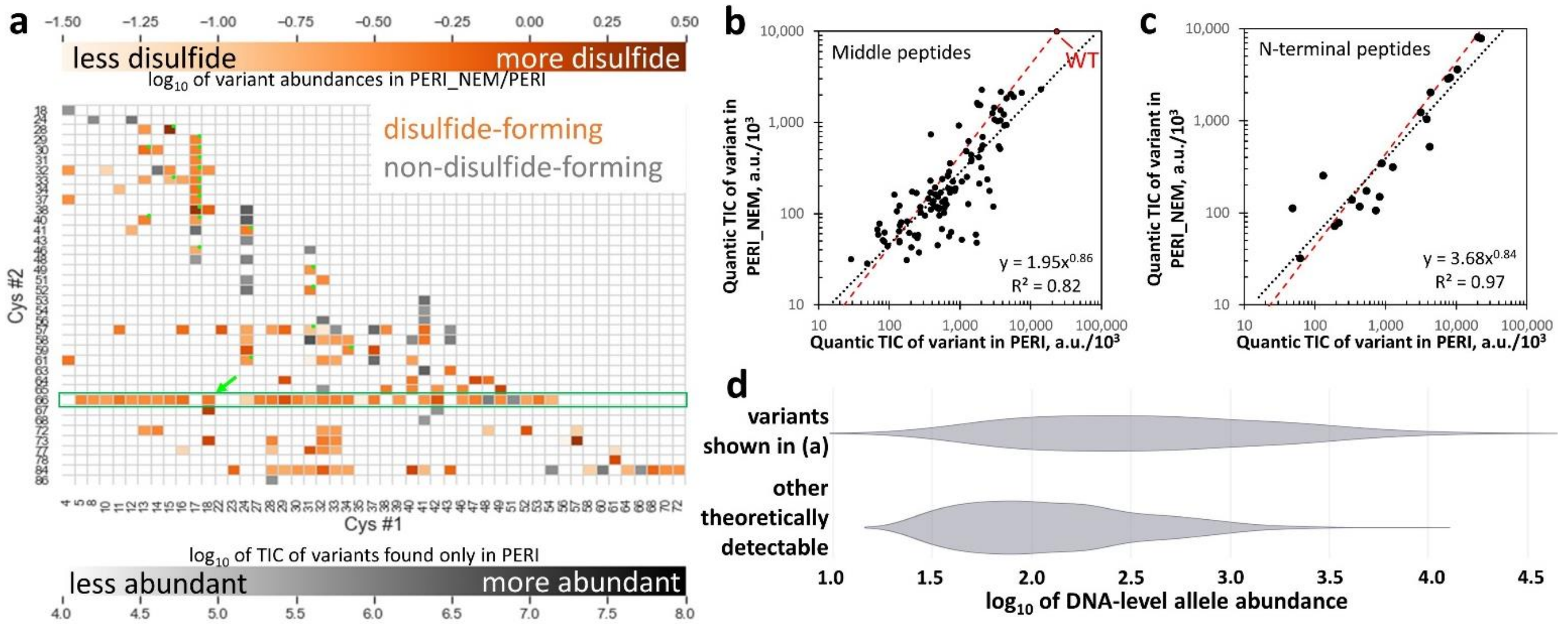

**Figure 4: Disulfide-bonding propensities of 166 double-Cys variants. (a)** Heatmap showing the abundance ratios (by total ion current) for variants found in NEM-treated ("PERI_NEM") and untreated ("PERI") periplasmic samples. Abundance of the most abundant variants found only without NEM is in gray. The *green arrow* indicates WT. Variants identified exclusively from N-terminal missed-cyanylation peptides are marked with *green dots*. The *green box* highlights variants containing Cys66, which were purposely enriched in the library. **(b)** Abundances in PERI and PERI_NEM (quantified by total ion current of the middle peptides, using Quantic) were strongly correlated, with a linear-regression trendline (*black dashed line*) close to the ratio measured for WT (*red dashed line*), so the vast majority of variants formed disulfides. Axes in thousands. **(c)** Same as **b** but for variants identified from N-terminal missed-cyanylation peptides only. **(d)** Violin plots of DNA-level allele abundance distributions (log-scale) of HdeA double-Cys variants detected vs. not detected at the protein level in the dataset in panel **(a)** illustrate that protein variants whose alleles were enriched at the DNA level in the library were much more likely to be detected in MS/MS. Only the theoretically detectable variants (Cys residues 11-62 peptide bonds apart) are considered.

## Absent or ectopic disulfides promote dissociation of folded HdeA dimers to disordered monomers.

To investigate the structural basis of toxicity, we used atomistic Monte-Carlo protein unfolding (MCPU) simulations with a knowledge-based potential; this freely available software was developed in our lab and shown to be effective in folding up small proteins[50] and predicting stability effects of mutations,[38] effects of non-native disulfides on aggregation,[39] and protein free energy landscapes[37] (version used here can be found at https://github.com/proteins247/dbfold). We chose 50 double-Cys variants plus noC, spanning various levels of toxicity and regions of sequence space, for multiplexed temperature-replica exchange simulations, using the WT homodimer (PDB ID 5WYO) as the starting structure. Intramolecular disulfides were modeled as strong flat-bottomed harmonic restraints.[39] Sampled structures from all simulated variants were clustered by similarity of intermolecular contact maps using DBSCAN.[51] This produced just three main clusters, plus numerous smaller clusters and non-clustering structures (**Figure 5a**). The largest cluster, termed "dissociated," comprised structures with few or no intermolecular contacts between the subunits. The second-largest, "C-term," comprised non-native dimeric structures, in which the subunits' C-termini unexpectedly bound each other as antiparallel helices. The third-largest, "WT-like," largely preserved the native homodimer interface. Most of the smaller clusters (963 of 1,029) had 28 or fewer structures each and thus could be unique conformations from a single coordinate replica of a single HdeA variant. A total of 7,484 structures were assigned as non-clustering.

The dissociated cluster grew larger at higher simulation temperatures (i.e., higher likelihoods of accepting energetically unfavorable steps[52]) (**Figure 5b**). Interestingly, dissociated and non-clustering structures were fewest at intermediate simulation temperatures where the C-term cluster was largest; we speculate that when both subunits lose their native structure, they may form the C-term structure before even higher temperatures melt it, too. Representative main-cluster structures are shown in **Figure 5c**. The WT-like structure of 32/66 shows that some non-native disulfides do not preclude native-like conformations, despite causing cytotoxicity. The many minor-cluster conformers in 32/66, however, may together constitute a molten globule. Many "C-term" structures, like the 40/65 structure shown, had helices 3 and 4 of a subunit aligned into a single, long helix, with the rest apparently molten; hydrophobic residues on helix 4, originally buried in the core, formed the new intersubunit interface. This kind of structure has not been experimentally observed, but it might explain some natively "forbidden" intermolecular crosslinks previously found *in vivo*.[53]

Strikingly, conformational changes from many non-native intramolecular disulfides propagated to the dimer interface, dissociating it (**Figure 5d**). Dissociation did not occur in noC under the same simulation conditions (**Figure 5d**, *"None"*). The sum total of dissociated, minor-cluster, or non-clustering conformers likewise often exceeded that of noC (**Figure 5d**). This counterintuitive observation of disulfides promoting disorder, together with the observation that disulfides formed *in vivo* whether or not they were consistent with the native state (**Figure 4a**), suggests that many double-Cys variants in **Figure 3b** may populate disulfide-trapped disordered conformations that are more cytotoxic than noC.

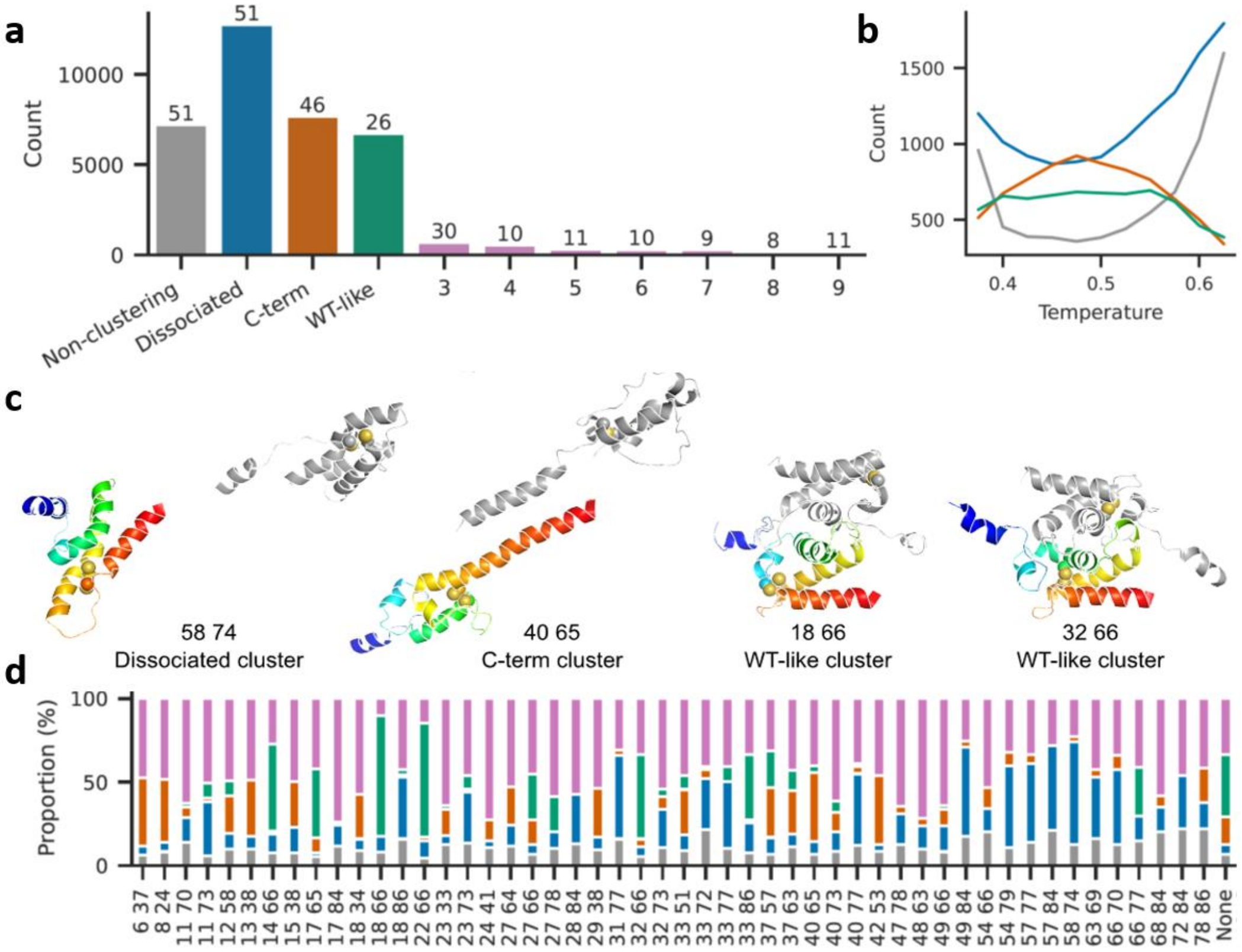

**Figure 5: Many non-native disulfides cause dissociation of HdeA dimers and melting of monomers in atomistic simulations. (a)** Combined sizes of top 10 clusters from all variant simulations; the number above each bar indicates how many variants had structures in that cluster. Total structures: 61600. Total clusters: 1032. Total structures belonging to none of the three main clusters: 29455. **(b)** Variation in cluster sizes with simulation temperature (colors as in **a**). **(c)** Representative structures from the three main clusters (numbers indicate Cys positions). The A subunit is in *rainbow colors* while the B subunit is gray. Cysteine residues are shown as spheres. Simulations included a 13-residue Gly-Ser linker between the subunits, here omitted for clarity. **(d)** Proportional cluster sizes for each simulated variant (colors as in **a**). "None" signifies the noC variant.

To experimentally validate these insights, we purified WT, noC, 32/66, and 40/65. The three mutants had quenched and red-shifted intrinsic tryptophan fluorescence spectra compared to WT (**Figure 6a**), indicating high solvent exposure of the two Trp residues (Trp16 and Trp 84). Gel filtration indicated similar hydrodynamic radii for WT and variants, with noC slightly more expanded and 32/66 and 40/65 slightly more compact. Yet, multiangle light scattering clearly indicated all three mutants were monomeric, unlike WT at the same concentration (**Figure 6b**). The WT's circular dichroism spectrum was consistent with its PDB structure (**Figure 6c**), but the mutants' spectra indicated much less helicity and predominantly random coil. This was consistent with the prediction of partial loss of native structure in **Figure 5d** but exceeded the predicted magnitude of structure loss. Although dimer dissociation was not predicted for 32/66 or noC, it was predicted for many other variants (**Figure 5d**). The need to connect subunits by a long linker in our simulations likely led us to overestimate dimer stability for 32/66 and noC; MCPU's statistical potential also does not fully incorporate the extra driving force for dimer dissociation due to HdeA's net negative charge (-4.1 per subunit for noC at pH 7).

Despite their very similar CD spectra, 32/66 was significantly more hydrophobic than noC (**Figure 6d**). This increased hydrophobicity was entirely attributable to the non-native disulfide because reduction eliminated it (**Figure 6e**). (WT's partially buried disulfide resisted reduction.) Hydrophobic interaction chromatography (HIC) easily separated 32/66 from WT due to the large difference in hydrophobicity (**Figure 6f**).

We additionally collected CD spectra of eight other double-Cys variants, of varying toxicity and from distinct regions of the sequence: 31/66 and 33/66 (to test whether a twist of the interface helix could rescue 32/66); 31/48 and 41/63 for comparison; apparent low-toxicity variants 12/58, 13/59, and 6/74; and near-neutral 17/68, whose Cys are close to the native 18/66 positions. CD, intrinsic fluorescence, and bisANS fluorescence spectra of all these variants (**Figure SI 14**) were similar to those in **Figure 6**. Thus, the native conformation appears to be highly fragile: it melts when the native disulfide is lost or shifted by even 1-2 sequence positions.

Having confirmed that non-native disulfides favored dissociated monomers, we then mapped predicted free energy landscapes using replica-exchange MCPU simulations of the monomers with umbrella biasing (by percent native contacts) and obtained unbiased statistics using MBAR in DBFOLD [54]. We focused on variants 32/66 and 18/32, both highly hydrophobic (**Figure 7**) yet differing in their toxicity (**Figure 3**), and noC and WT monomers for reference. Fraction of native contacts *Q* and $C_\alpha$ root-mean squared deviation (RMSD) were defined with respect to equilibrated WT monomer. Free energy landscapes (**Figure 6g**) showed a single native-like basin for WT but two basins for noC: native-like and non-native. The non-native basin was much more populated in 32/66 variant than in 18/32 along the RMSD coordinate (**Figure 6i**), though not along the Q coordinate (**Figure 6h**). These findings suggest that 32/66 favors a non-native cytotoxic conformational basin already present in the underlying free energy landscape. Average equilibrium surface hydrophobicity of monomeric noC, 32/66, and 18/32 was higher than that of dimeric WT simulated the same way (**Figure 6j**), consistent with the experimental trend (**Figure 6d,e**).

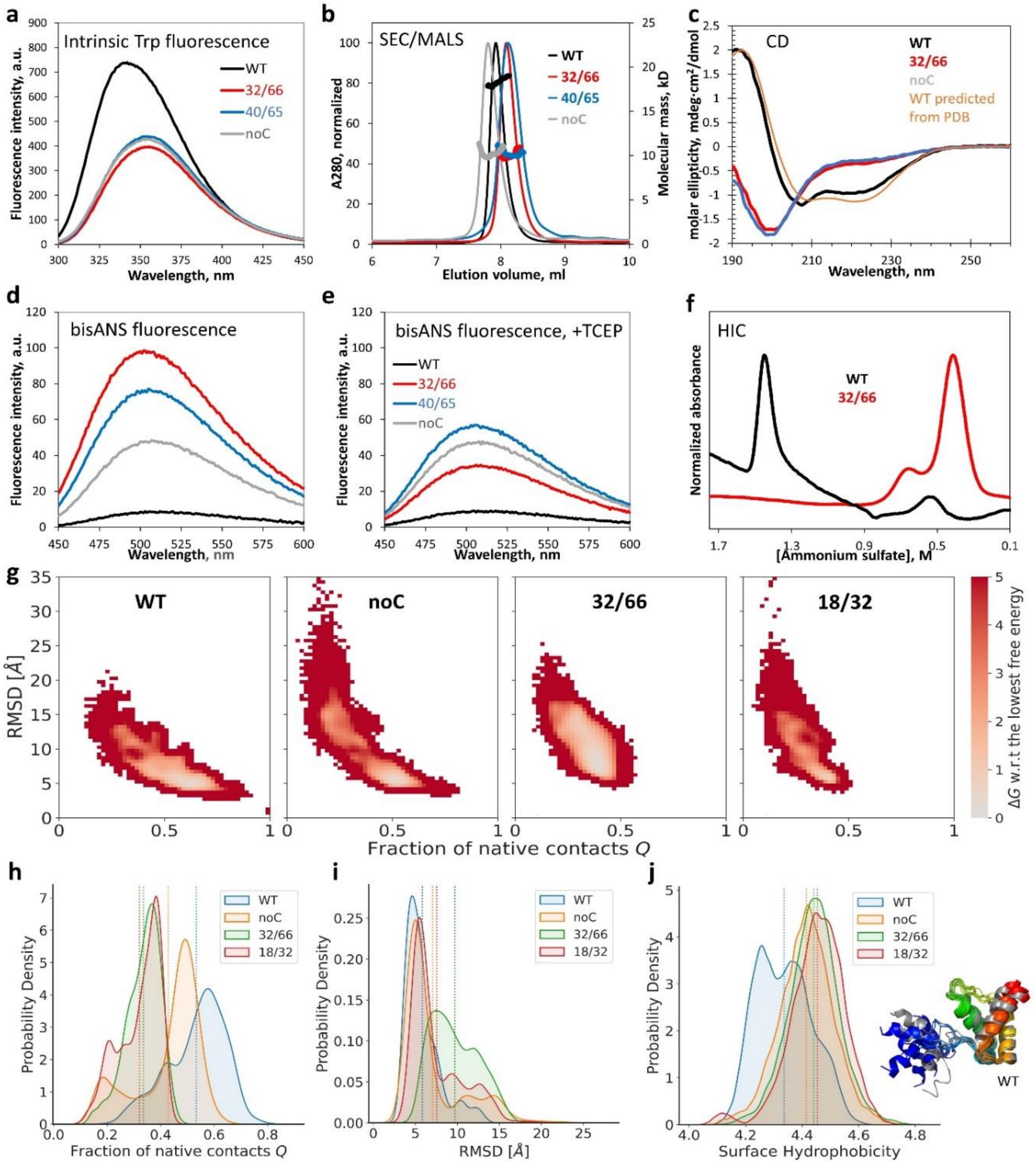

**Figure 6: Biophysical characterization indicates HdeA mutants are monomeric, highly disordered, and hydrophobic. (a)** Intrinsic tryptophan fluorescence was quenched and red-shifted in 32/66, noC, and 40/65 compared to WT. **(b)** SEC/MALS revealed that, despite similar elution positions, only WT was dimeric; all mutants were monomers. **(c)** CD of WT matched predictions from PDB ID 5WYO by the PDB2CD tool (https://pdb2cd.cryst.bbk.ac.uk), but 32/66 and noC clearly did not. **(d)** The hydrophobicity probe bisANS bound much more strongly to the mutants (same samples as in **a**), especially 32/66. **(e)** Upon reduction of the disulfides, 32/66 and 40/65 bisANS fluorescence resembled noC; WT was resistant to reduction. **(f)** Folded (WT) and disordered (32/66) were easily separated by HIC. (**g**) Calculated free energy landscapes of monomeric WT, noC, and two double-Cys variants from MCPU simulations; the monomeric WT starting structure is defined as Q = 1. (**h**) The distribution of Q-values (fraction of native contacts) for mutant and WT HdeA. (**i**) The corresponding distribution of RMSD values relative to the starting WT structure. (**j**) The distribution of surface hydrophobicity values for WT dimer and mutant monomers; although it appears bimodal for WT, representative structures of the WT subunits (*rainbow- vs. gray-colored ribbon diagrams*) did not appreciably differ outside the N-terminal disordered region (*blue/gray*), which was often helical in our simulations. *Dashed lines* in panels **h-j** are distribution averages.

## Allele toxicity correlates with protein hydrophobicity among core HdeA variants.

To investigate how allele toxicity relates to biophysical properties, we fractionated the protein variants by HIC, where hydrophilic proteins elute first (i.e., at higher [ammonium sulfate]) and hydrophobic ones later. This protein library was purified from culture supernatants as the more abundant protein source: 339 protein variants were detected in at least one HIC fraction (**Figure 7a**). We found excellent batch/batch agreement in protein variant quantification (**Figure SI 15**).

Due to the wide range of variant signal intensities relative to MS/MS dynamic range (see above), just 39 variants were detected across at least four HIC fractions each, yielding interpretable elution profiles. Of these, 17 were "core" variants, i.e., neither Cys residue was in the N-terminal flexible region (defined here as residues 1-17). They were identified mostly from middle peptides, and most of the others from N-terminal peptides. We clustered the core variants into four types of HIC elution profiles, defined by the centroid of abundances across the fractions where the given variant was found. WT and 70/84 were the least hydrophobic in this set and the least toxic (**Figure 7b**). Moderately hydrophobic variants (**Figure 7c**) were typically moderately toxic. More hydrophobic variants (**Figure 7d**) were also more toxic, except 18/32, which resembled the "N-terminal" variants in **Figure 7f**: hydrophobic proteins yet only modestly toxic alleles. The most toxic alleles produced proteins with anomalous saddle-shaped elution profiles (**Figure 7e**), suggesting transient burial of their hydrophobics, either via folding or (more likely) via interaction with other proteins or even transient aggregation. **Table SI 2** lists the hydrophobicity centroids and measured allele toxicities of the 39 variants in **Figure 7**.

Most variants in **Figure 7** were also detected in **Figure 4** and confirmed to form disulfides. The exceptions were 37/61 and 31/58; these non-disulfide-forming were also mildly-hydrophobic and (**Figure 4a**) and mildly toxic – resembling noC. Unlike 31/58, variant 31/57 formed the disulfide some of the time (**Figure 4a**) and was more hydrophobic and more toxic (**Figure 3b**); nearby variant 29/57 showed even more disulfide bonding, greater hydrophobicity, and higher toxicity still. Anecdotally, then, core variants with noC-like phenotypes had noC-like structures (i.e., no disulfide). The only hydrophilic mutant, 70/84, was disulfide-forming yet had only mild (noC-like) allele toxicity. All the "N-terminal" variants in **Figure 7f** also formed disulfides.

The PSM-weighted average allele toxicity (*E*) of each HIC fraction showed a strong, apparently sigmoidal correlation with hydrophobicity for the 235 variants identified from middle peptides but no such correlation for the 104 variants identified from N-terminal peptides only (**Figure 7g**). The difference is likely due to enrichment of core variants in the former set of peptides and N-terminal ones in the latter. Indeed, plotting hydrophobicity vs. allele toxicity of the 39 most-abundant variants, without any PSM-weighting, revealed a statistically significant ($p = 0.02$) correlation for the core variants, identified from either type of peptide (**Figure 7h**), which became even stronger ($p = 0.003$) after correcting for centroid shifts due to the saddle-shaped HIC profiles of extremely toxic variants (**Figure 7i**). The non-core variants showed no such correlation (**Figure 7j**). Thus, protein hydrophobicity was necessary but not sufficient for high allele toxicity: N-terminal variants showed the former but not the latter.

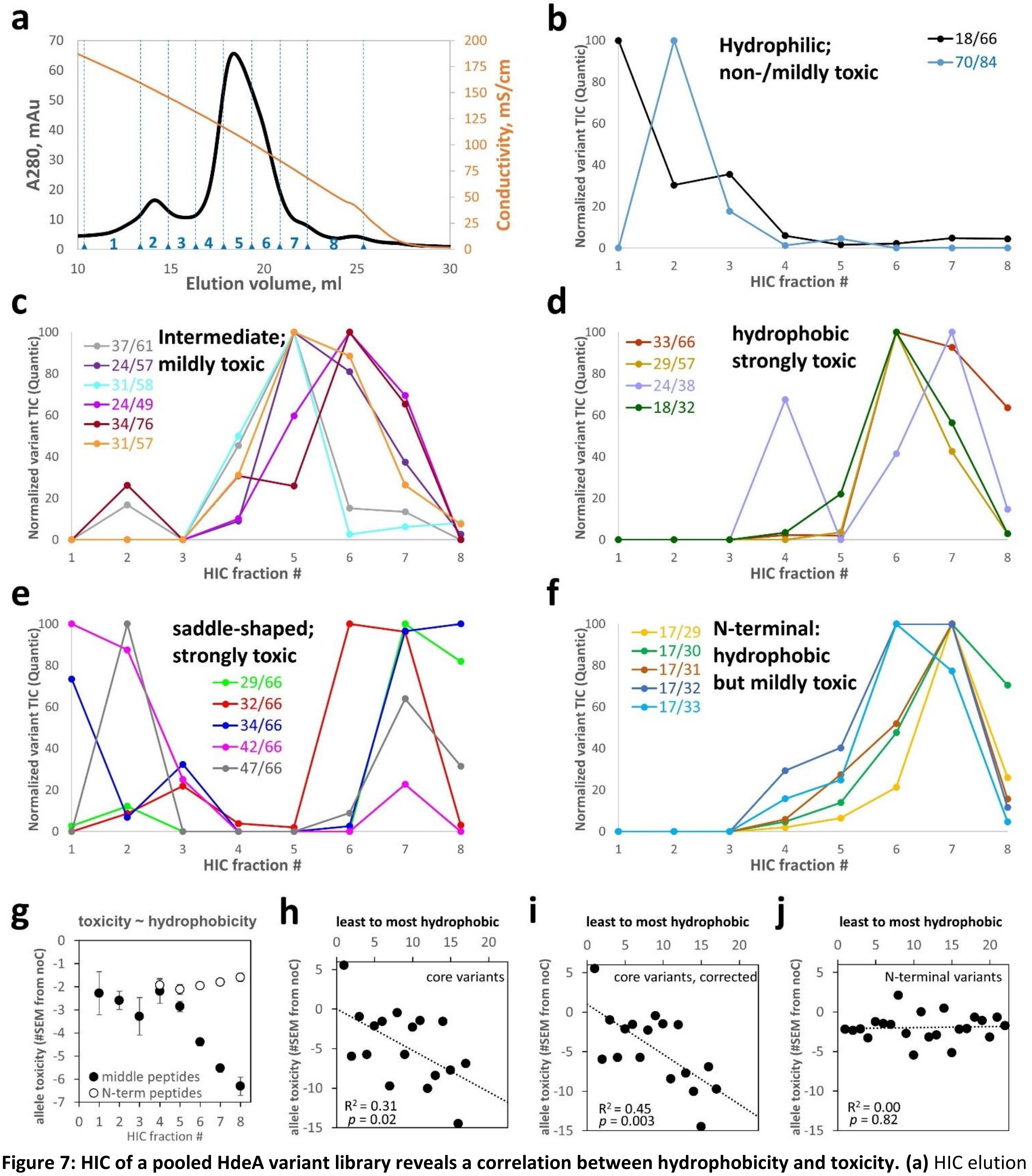

**Figure 7: HIC of a pooled HdeA variant library reveals a correlation between hydrophobicity and toxicity. (a)** HIC elution trace of the total HdeA variant library on an ammonium sulfate gradient reported by conductivity (*orange curve*). Eight fractions were collected, numbered as indicated. Elution profiles of all 17 "core" variants are shown, grouped by similarity of elution profiles and centroids of elution: **(b)** below 4.7, **(c)** 4.7-6.0, **(d)** 6.0-6.8. **(e)** Some toxic core variants had unusual saddle-shaped elution profiles. **(f)** Most non-core variants were identified from N-terminal peptides, many with the missed cyanylation at position 17; those variants had high hydrophobicity despite modest toxicity. **(g)** Hydrophobicity and toxicity (PSM-averaged *E* value) were strongly correlated for variants identified from middle peptides but not from N-terminal peptides. **(h)** Differences between core and N-terminal variants accounted for this divergence, especially after correcting for centroid shifts due to saddle-shaped elution profiles (by ignoring data from the first 3 HIC fractions) **(i). (j)** By contrast, non-core variants – identified from either set of peptides – showed no hydrophobicity-toxicity correlation.

## Discussion

We have demonstrated the feasibility of disulfide-scanning an entire protein in one experiment: at the DNA level using amplicon deep-sequencing and at the protein level by devising and implementing a novel method for limited deep-sequencing of proteins. High-throughput disulfide scanning (HTDS) can map conformational landscapes to molecular and organismal phenotypes in a model-free manner. Our approach is widely accessible: the total cost of all DNA and protein deep-sequencing reported in this study was just over $1,000. HTDS will be even more powerful when longer-read single-molecule protein deep-sequencing becomes available.

We found direct experimental evidence for misfolding toxicity in HdeA, a conditional IDP and one of the most abundant *E. coli* proteins. Removing the native 18-66 disulfide was toxic but shifting it by even one residue (to 18-67 or 18-65) or one helical turn (22-66 or 18-70) was even more toxic. Only a few near-native disulfides (17/66, 17/65, 15/66) significantly (>2 SEM) rescued noC toxicity. This brittleness is surprising given the high thermostability of the folded WT structure.[55] HdeA may have evolved brittleness because function requires on-cue denaturation. HdeA's highly disordered conformational ensemble in the absence of its native disulfide explains why most of the detected double-Cys variants were capable of intramolecular disulfide bonding (**Figure 4a**). Ectopic disulfides like 32/66 raised both hydrophobicity and cytotoxicity above noC, and reducing them restored noC-like hydrophobicity. Thus, some disulfides may trap otherwise transient cytotoxic conformers on a landscape resembling **Figure 1a**. Sharp phenotypic peaks on the backbone conformational landscape may be common in IDPs: e.g., a recent study found that the 17/28 internal disulfide in amyloid-$\beta_{42}$ led to formation of highly cytotoxic oligomers, but 16/29 or 18/27 did not.[56]

The detailed molecular mechanism of HdeA toxicity remains a topic for future research. We have shown that it entails protein misfolding stress, evidenced by strong DnaK overexpression, and ultimately cell lysis. Protein misfolding is frequently cytotoxic; proposed explanations include direct disruption of cell membranes, aberrant protein-protein interactions, or overloading of protein quality control systems.[15,57] Elucidating HdeA toxicity mechanisms could uncover new bacterial physiology, especially given our unexpected observation that WT HdeA is efficiently exported without rupturing the cell. It is also surprising that DnaK overexpression should be the most obvious cellular response, given that DnaK is in the cytoplasm and HdeA in the periplasm. However, there is strong proteomic evidence that DnaK natively interacts with HdeA and stabilizes it, and both HdeA and HdeB are degraded in Δ*dnaK* cells.[58] DnaK overexpression has also been observed to increase periplasmic expression levels of some exogenous proteins, perhaps by enhancing the export of these proteins from the cytoplasm to the periplasm.[47] To the best of our knowledge, the mechanism of stabilization of HdeA by DnaK is not understood.

This study's main technical innovation is protein-level deep-sequencing of double-Cys protein variant libraries to identify their redox state or chromatographic profiles. Both these applications are illustrated in our study. The former could make HTDS a powerful approach for *in vivo* structural biology. Low-throughput disulfide scanning can already distinguish structural models or clarify specific tertiary structure elements.[19-22] We expect that HTDS will allow model-free structure determination, at moderate resolution but *in vivo*, including for non-native and perhaps even aggregated conformers. If many IDPs are indeed ordered *in vivo*,[16] HTDS could help elucidate those structures. Thanks to the latter application, HTDS can go still further and enable many types of biophysical measurements, such as thermodynamic stability, binding affinity to a target, or aggregation propensity, at much higher throughput than conventionally possible. When combined with DNA-level deep sequencing of the variant libraries, this method can enable full pipelines of genotype-conformation-phenotype analysis.

Importantly, HTDS can not only reveal native or non-native conformations but also stabilize them. This is critical for biomedical applications like vaccine design or drug screening. Stabilizing the wrong conformation of the antigen in the RSV vaccine led to tragic consequences.[59] Low-throughput disulfide engineering later enabled stabilization of the correct conformation.[60] The same approach has been applied to the SARS-CoV-2 spike protein, resulting in some useful prefusion stabilization[61,62] but also a very high failure rate.[63] In those studies, the sites for disulfide engineering were chosen by some combination of structural intuition and computational modeling, and <100 variants were screened. HTDS could allow screening of many thousands of variants even before an atomistic structure is available and may reveal useful disulfides with unexpected allosteric effects. There are also potential applications in the treatment of neurodegenerative and other conformational diseases. Misfolded aggregation precursors stabilized via disulfide crosslinking could become protein vaccines. There is encouraging biological precedent in the discovery that some seniors are protected from ALS because their immune systems have generated conformation-specific antibodies that clear the misfolded aggregation precursor.[64] Finally, identifying disulfides that trap a non-native cytotoxic conformation – as we report here for *E. coli* HdeA – could enable screening for drugs that bind and stabilize that conformation even in the wild-type protein. Such compounds could become a novel class of antibiotics, which would work by induced misfolding of target proteins.

## Limitations

Library coverage was lower for proteins than DNA, mainly due to the much lower practical dynamic range of MS/MS compared to Illumina sequencing. Allele abundances varied by >1000X in our plasmid library – partly because of the exponential PCR amplification steps of megaprimer-based mutagenesis and partly by design, as the Cys66-containing sub-library was enriched at the DNA level. Library coverage was clearly better for the enriched variants (**Figure 4a**), with 31 variants detected by LC/MS/MS out of the 56 detected by Illumina. As high-throughput DNA synthesis advances, *de novo* synthesis of double-Cys libraries of desired composition (e.g., 100% double-Cys variants instead of ~47% in this study) and more uniform allele abundances will improve protein-level coverage.

The maximum length of peptides amenable to MS/MS analysis presents another major current limitation. In this study, we were limited to peptides of 12-63 residues total, not counting the N-terminal ITZ group but counting the C-terminal incorporated nucleophile (Arg). The lower bound of this range is somewhat arbitrary but useful to avoid unrealistically high apparent confidence in PSMs due to very short peptides. The upper limit is partly software-driven, since Comet currently cannot handle peptides longer than 63 amino acids; however, it is also partly hardware-driven, as longer peptides tend to fragment poorly in MS/MS and yield fragments of very high and variable charge state.

A third important limitation is that the current version of HTDS works best for natively Cys-less proteins. It is possible to apply it to proteins with non-bonded native Cys residues (by alkylating them before protein-level deep sequencing) or even to proteins with native disulfide bonds, but only if the scanned region of the protein does not overlap the native disulfide. If it does, then the crucial "middle" peptide in **Figure 1c** will be cleaved during the aminolysis step, thus severing the physical linkage between the two original Cys positions and abrogating the ability to identify any double-Cys variants whose sequence positions overlap the native disulfide.

All three of the limitations above will be fully eliminated when single-molecule long-read deep sequencing of proteins (at least with respect to Cys locations) is developed. Our current method based on cyanylation-aminolysis and MS/MS is a bridge technology in this regard. HTDS should be a motivating application for the invention of MS-free methods, such as those based on nanopores,[65] massively parallel Edman-based[66] or terminal peptidase-based[67] sequencing, magnetic tweezers,[68] DNA nanoscopy,[69] or hyper-resolution fluorescence microscopy.[70]

Lastly, it is informative to contrast our HTDS method with chemical crosslinking MS (XL/MS), which can also probe protein structure when combined with *in-silico* simulations.[71,72] HTDS has distinct advantages and limitations relative to XL/MS for structure determination. Disulfides easily form *in vivo* without external perturbation, even in cytoplasmic proteins in strains such as SHuffle®.[73] Being genetically encoded, disulfides can reveal heritable and evolutionary effects of protein conformations. HTDS as implemented here reports only intramolecular crosslinks, unambiguously identified from the termini of linear peptides, which greatly simplifies analysis. Finally, the disulfide bond's combination of small size and reversibility is unmatched by chemical crosslinkers. On the other hand, HTDS cannot be applied to entire proteomes, as limited proteolysis/MS has already been applied[16] and XL/MS could be, in principle. An oxidizing environment is required, and some mutations could interfere with folding. Finally, HTDS in its current form cannot be applied to natively disulfide-rich proteins; only full single-molecule protein sequencing, without cleavage, will likely overcome this limitation. Finally, XL/MS is better for mapping quaternary structure.

# STAR Methods

*Generation of 2-Cys scanning libraries*

A linear dsDNA fragment encoding the C18S/C66S variant of *E. coli* HdeA ("noC"), with the native signal sequence for periplasmic export, was purchased from GeneUniversal. Except for the two point mutations, the sequence was entirely native, with no codon optimization. This fragment was cloned into the pCK302 vector[46] in place of the superfolder GFP gene, using the FastCloning method.[74] The vector was obtained via Addgene. In subsequent validation experiments, all indicated individual variants were synthesized by GeneUniversal or Thermo Scientific and cloned into the same vector in the same way. A single-Cys scanning library covering all 83 positions from Gln4 to Lys86 (numbered based on the mature HdeA protein, without signal sequence) was purchased from GeneUniversal and cloned into the pCK302 vector using FastCloning. Limited T5 exonuclease digestion[75] was used to increase cloning efficiency, resulting in ~200 transformant clones.

Mutagenesis of the single-Cys scanning library ("CSL") was carried out using the megaprimer method.[76] First, short mutagenic forward primers, covering 51 positions from Ala3 to Lys87, were purchased from IDT DNA. These 51 positions were chosen to avoid mutations to Pro, Gly, Trp, Phe, Leu, Ile, Glu, or Asp codons at this stage. The noC gene was PCR-amplified in 51 separate reactions using those forward primers and a fixed reverse primer downstream of the gene, in addition to a primer-less negative control. The primer sequences are listed in **Table SI 1**. The resulting megaprimers were directly applied, without further purification, to a new set of 51 PCR reactions, this time with CSL as the template instead of noC. The pooled PCR products were incubated at a 10:1 ratio with DpnI enzyme (New England Biolabs) at room temperature for 5 min, then at 37 °C for 50 min. Self-ligation of the long linear PCR fragments was achieved by *in vivo* homologous recombination, assisted by limited T5 exonuclease digestion, as above, prior to transformation to DH5α competent *E. coli* cells (New England Biolabs), to create a multi-Cys scanning library ("MCSL") in the pCK302 vector. Separate digestions and transformations were carried out for targeted libraries having either or the two native Cys residues, Cys18 or Cys66, on the CSL background (termed "C18CSL" and "C66CSL"), as well as for the negative-control library. After recovery in SOC medium for 1.5 h, samples representing 2.5% of the volume of each library were plated on LB-agar plates containing ampicillin, resulting in 200-300 colonies each for MCSL, C18CSL, and C66CSL, and no colonies for the negative control, indicating ~10,000 transformants per library. The remaining 97.5% of each transformation mixture was inoculated directly into SuperBroth (Teknova) containing ampicillin, to a total volume of 4 ml, cultured overnight at 37 °C with shaking. Plasmids were prepared using the Qiagen miniprep kit with two preps per culture.

A total of 10 individual clones from MCSL were Sanger-sequenced, along with two clones each from the targeted CSLs. The MCSL clones included two 1-Cys variants, six 2-Cys variants, and two 3-Cys variants; the C66CSL clones were two 2-Cys variants; and the C18CSL clones were one 1-Cys variant (containing C18) and one variant with a run-on duplication of the primer.

For all further experiments, a combined library ("MCSL++") was generated to enrich for the native Cys residues and ensure reasonable abundance of variants that could serve as markers for the DNA-level and protein-level sequencing of the pooled library cultures. Specifically, 830 ng of the MCSL library was combined with 130 ng of the C66CSL library and 10 ng each of the cytotoxic 40/65 variant and the non-toxic WT.

*Allele toxicity assays*

All toxicity assays were carried out in the non-rhamnose-metabolizing *E. coli* BW25113-derived Δ*hdea* strain from the Keio knockout collection. The strain was plated, and one clone selected for further work. The cells were made chemically competent by a modified version of manganese-assisted permeabilization.[77] Filtered, chilled Inoue solution was prepared from 25 ml MilliQ water, 0.5 g KCl, 0.25 g $MnCl_2 \cdot 4H2O$, 0.055g $CaCl_2 \cdot 4H2O$, and with 10 mM (final) PIPES buffer pH 6.7. Two cell pellets grow in SuperBroth medium (Teknova) were washed with 8 ml of this solution each, then resuspended in a combined 4 ml volume of the same solution, always on ice. Room-temperature DMSO (0.3 ml) was then added, along with 1.7 ml 50% glycerol if cells were to be stored at -70 °C. Otherwise, the cells were transformed immediately. All constructs used in the allele toxicity assay were then transformed into this strain of competent cells. In the case of MCSL++, 1% of the transformant mixture was plated, resulting in 96 colonies. The full library therefore contained ~10,000 transformants, which was deemed sufficient.

Thawed polyclonal glycerol stocks were diluted 1:100 to fresh LB broth with 100 µg/ml ampicillin and 50 µg/ml kanamycin and allowed to recover for 1-2 h at 37 °C with shaking. Induction cultures in the same medium containing various amounts of L-rhamnose in 96-well plate format were then inoculated 1:100 or 1:1000 (as indicated) from these recovered cultures. The 96-well plates were incubated for 14 h at 37 °C with 567 cpm linear shaking (3 mm amplitude) in BioTek Epoch plate readers with optical density monitored at 600 nm.

*Amplicon library preparation and deep sequencing*

Replicate time point samples from pooled cultures of the MCSL++ transformant library were centrifuged for 15 minutes at 3,000*g* in 15-ml conical tubes or 14-ml round-bottom tubes to fully clarify them. The supernatants were then carefully withdrawn by pipetting and both pellets and supernatants iced or frozen thereafter. Plasmids were extracted from the pelleted cells using a GeneJet (Thermo) miniprep kit according to the manufacturer's instructions. Amplicons with Nextera adapters for Illumina sequencing were generated from the pelleted cells ("PEL") using 10 ng of purified plasmid as the template for each 25 µl reaction and 25 PCR cycles with the Q5 Master Mix (New England Biolabs). Amplicons from the clarified culture supernatant ("SUP") were generated the same way, using 1 µl of the SUP sample as the template. PCR cleanup was carried out on each amplicon sample using the Monarch PCR cleanup kit (New England Biolabs). The amplicons were then dual-indexed for paired-end read Illumina sequencing using Nextera primers N701 and N702 for the PEL and SUP of the 5.5 h time point and N705 and N706 for the PEL and SUP of the 9.0 h time point, respectively, and primers S517, S502, S503, and S504 for samples from the four replicate cultures in each case. Per manufacturer instructions, 8 PCR cycles were used, with 100 ng template per 50 µl reaction with Q5 Master Mix. Success of each reaction was confirmed by running 5 µl of each indexed PCR produced on an agarose gel with ethidium bromide staining, which showed bands of equal intensity for all. The samples were then pooled into 8-plex libraries by time point (containing equal volumes of PCR products for all replicates of both PEL and SUP amplicons), each pool PCR-cleaned again as above, and the resulting indexed 8-plex amplicon libraries sequenced by Novogene in a single split NovaSeq lane using paired-end 155 base-pair reads. The reads were merged using AmpliMERGE[78] and the reads filtered, trimmed, and variants counted (with all synonymous reads combined) using Enrich2.[79] For the 9.0 h time point, 1.8-2.2 million reads per replicate were identified, and for the 5.5 h time point, 1.2-1.7 million reads per time point. The 5.5 h data were not used further due to very high variability: a large number of outliers, particularly in the SUP sample, with one of the four replicates showing >10X higher abundance than the others. This may be attributable to the very low amount of DNA in the culture supernatants at this early time point. Therefore, we focused on the 9.0 h time point, where such cases were few. Only variants with at least 50 reads on average across all conditions were included in further analysis. Variants containing outliers were filtered out by requiring that the standard error of the SUP/PEL abundance ratio be no more than half as large as the ratio itself. The SUP/PEL ratio was used as the raw measure of allele toxicity, since protein variants that caused cell lysis were the ones whose plasmids were disproportionately enriched in the supernatant of the pooled cell cultures.

*Calculation of epistatic allele toxicity of the 2-Cys variants*

Ratio epistatic toxicity [48] for each double-Cys variant was calculated by dividing its SUP/PEL abundance ratio by the product of the SUP/PEL ratios for the two single-Cys variants at those positions. For the noC variant, toxicity was defined as just the raw SUP/PEL ratio, which was 0.89 for the 9.0 h time point samples. Since variant abundances in the MCSL++ library varied by at least three orders of magnitude at the DNA level (partly by design), the standard errors also varied, tending to be greater for the less-abundant variants. To account for this, we chose to plot the allele toxicity landscape of **Figure 3b** in terms of the number of standard errors by which the mean of epistatic toxicity differed from the mean of noC (i.e., from 0.89). The resulting map did not differ qualitatively from the more traditional measure of the logarithm of ratio epistasis (shown in Figure **SI 7**). Heatmaps were generated using Matplotlib[80] and Seaborn[81] in Python 3.

*Low-throughput verification of HdeA variant expression levels*

We used a random number generator (www.random.org) to select a total of 40 variants, 20 each from the top 10% ("non-toxic") and bottom 10% ("toxic") of the epistatic allele toxicity (#SEM from noC) distribution, as in **Figure 3b**. The chosen sequences were then synthesized, cloned into the same rhamnose-inducible pCK expression vector and expressed in the same strain of *ΔhdeA E. coli* as in all prior experiment. Two variants (both "non-toxic") were discarded due to sequence errors, leaving 38 variants. Polyclonal glycerol stocks were prepared from 2-ml LB/Amp/Kan cultures inoculated with 5 colonies picked for each variant and grown for 4 h at 37 °C and 250 rpm orbital shaking in 14-ml tubes. Control cultures of WT, noC, 32/66, 18/86, and sGFP were prepared at the same time. In three independent replicate experiments, passaging cultures were inoculated from thawed glycerol stocks at 1:10-1:100 ratios and grown in a sealed 200-µl 96-well plate for ~2 h in LB/Amp/Kan, then inoculated 1:100 into a fresh identical 96-well plate with added 1 mM (final) L-rhamnose as inducer. In a fourth independent experiment, the induction plate was inoculated directly from glycerol stocks at 1:100 ratio. To measure expression levels, total culture samples for non-reducing SDS-PAGE were collected from the end-point of the 14-h induction assay in each case. One set of samples came from the passaged plate and the other from the directly inoculated. Gels were stained with GelCode Blue Safe Stain (ThermoFisher) and quantified using GelAnalyzer software (with default rolling-ball background subtraction), exactly as those in **Figure 2**. Both [DnaK] and [HdeA] measurements were internally normalized to the most intense band in each lane (~37 kD). The magnitude of growth curve dips was determined from $OD_{600}$ at time = $t_{lag}$ + 5 h as the % decrease from the corresponding value for WT (with any curves rising above WT assigned as a growth curve dip of 0).

*Protein expression*

Small-scale (typically 5 ml) LB cultures 100 µg/ml ampicillin and 50 µg/ml kanamycin were inoculated 1:100 with thawed glycerol stocks of the transformed *E. coli* BW25113 Δ*hdea* and allowed to recover for 1-2 h. In the case of MCSL++ cultures, the initial inoculate was 1:1000. Larger batches (0.5 L) of the same medium were then inoculated 1:1000 from these recovery cultures and incubated in 2L flasks at 37 °C with 250-300 rpm shaking for 14-16 h in the presence of 1-3 mM L-rhamnose as the inducer. Protein expression was verified by SDS-PAGE of both total and supernatant fractions of samples taken immediately thereafter. Cultures were harvested in conical or square bottles by centrifugation in a swing-bucket rotor. The supernatants were further centrifuged in conical bottles (Celltreat) at maximum speed, treated with Complete EDTA-free protease inhibitor mixtures (1 crushed tab per 1-2 L medium), and passed through bottle-top 0.2-µm filters (VWR). Thus treated, culture supernatants were stored tightly capped at 4 °C in sterile bottles until further use. As long as the cultures remained clear, SDS-PAGE did not show any noticeable protein degradation even after several weeks.

*Preparation of periplasmic extracts*

Extraction buffer comprised 330 mM Tris pH 8, 1 M sucrose, and 2 mM EDTA. Immediately prior to extraction, one crushed tab of Complete EDTA-free protease inhibitor cocktail (Roche) was added to ice-cold extraction buffer. Pelleted cell cultures were gently resuspended on ice in this buffer using a cell scraper and allowed to incubate for 10 or 30 minutes (the shorter time reduced the amount of complete cell lysis when toxic variants were expressed and was therefore used for the MCSL++ library). The extracts were immediately clarified by centrifugation at 12,000 rpm in a microcentrifuge chilled in advance to 4 °C. Extracts were then stored at -70 °C.

*Protein purification*

Since the majority of overexpressed HdeA was typically found in the culture supernatant (see **Figure 2b** and **Figure SI 5**), all protein samples except for the periplasmic samples in **Figure 4** were purified from supernatants prepared as described above and concentrated ~50-100-fold in Centricon Plus70 concentrators (Millipore). The concentrated supernatants were filtered through 2x5ml tandem Sepharose Q columns (Cytiva) in pH 6.7 10 mM PIPES buffer; mixed 1:1 with 4 M ammonium sulfate (Teknova) and centrifuged for 10 min. at 18,000g, then fractionated by hydrophobicity on a HiTrap Phenyl HP column (Cytiva) with 2-0 M ammonium sulfate gradient in 10 mM PIPES pH 6.7 buffer. Finally, the peak HIC fractions were concentrated as needed and fractionated by size exclusion on a Superdex 75 Increase 10x300 column (Cytiva) equilibrated in buffer suitable for the given experiment (10 mM sodium phosphates pH 7 for CD; the same with 150 mM NaCl for SEC/MALS; or 10 mM PIPES pH 6.7 with 150 mM NaCl for cyanylation/aminolysis).

Purification of the pooled libraries followed the same procedure, but with a step gradient in HIC elution (from 2M, then 1.5M, then 0M ammonium sulfate), due to the varying hydrophobicity of library components. This procedure resulted in ~90% purity for the MCSL++ library as determined by SDS-PAGE, which was deemed an acceptable compromise given the ability of mass spectrometry to identify peptides in complex mixtures.

Purification of periplasmic extracts was carried out exactly as above, except that NaCl was added to the periplasmic extracts to 150 mM final concentration prior to the Sepharose Q filtration step to minimize unwanted binding to the column.

*Cys-specific cleavage of proteins by cyanylation/aminolysis*

CDAP (Sigma) was prepared as a 100 mM stock in pure acetonitrile, the headspace of the vial purged with nitrogen, the vial lid taped, and the vial stored at -20 °C. Before being opened for use, the vial was allowed to equilibrate to room temperature to minimize condensation of water vapor from the air, and after use it was immediately purged, taped, and stored as before. We found that when this procedure was carefully followed the reagent retained most of its activity for multiple uses over the course of several months.

A near-saturated solution of 6 M L-Arginine at pH 9 was prepared by dissolving the powdered compound stepwise in a minimum amount of water while titrating the pH at each step with sodium hydroxide pellets and neat hydrochloric acid to ensure solubility. The solution was stored at room temperature protected from light and remained usable for several months.

Protein samples (10-100 μM concentration) were denatured in 4 M guanidinium chloride (Thermo), pH7, at 42 °C for 30-60 min. in LoBind microcentrifuge tubes (Eppendorf). For experiments identifying disulfide bonding propensities, this incubation was carried out in the presence of either 5 mM TCEP (for the total sample) or 10 mM N-ethyl maleimide (NEM, for the free thiol-blocked sample), with 10% v/v DMSO in each case (since N-ethyl maleimide stock was prepared in DMSO). Both samples were then fractionated by SEC on a Superdex75 Increase 10x300 column equilibrated in pH 7 sodium phosphates with 4 M guanidinium chloride, and the elution peaks (at ~10.0 ml for the TCEP-treated sample and ~10.4 ml for the NEM-treated sample) collected manually. Note that under the same conditions fully reduced WT HdeA eluted at 10.0 ml and non-reduced WT at 10.6 ml. This denaturing SEC step was expected to remove any disulfide-bridged dimers from the NEM-treated library. Prior to cyanylation, the SEC peak fraction of the NEM-treated library was reduced with TCEP for 45 min at 37 °C. For the proof-of-concept experiments on hydrophobicity, no SEC or thiol blocking was carried out to minimize sample loss. Instead, samples were reduced for 1 h with TCEP as above, then buffer-exchanged by Zeba desalting columns directly into the cyanylation buffer.

For cyanylation, all samples were buffer-exchanged to pH 3 buffer containing 20 mM sodium citrate and 4 M guanidinium chloride, either by ultrafiltration (using Pall Nanosep filters spun at 6000 rpm in a microcentrifuge) or by centrifugal desalting columns (ThermoFisher Zeba 2 ml) and immediately used for cyanylation without the addition of more reducing agent. To cyanylate the Cys residues, CDAP reagent was added to 10 mM final concentration, and the mixture was incubated at room temperature for ~2 h. Each sample was then mixed 1:1 with the 6 M pH 9 L-arginine solution and incubated at room temperature overnight, protected from light. The aminolysis reaction was then quenched by addition of 2% v/v of 95% formic acid, bringing the mixture to pH ~4.

*Identification of pooled 2-Cys variants by mass spectrometry*

A total of ~1 nmol of each cleaved sample was injected on a C18 HPLC column and fractionated with a 120-min gradient of 10-45% acetonitrile, followed by a 10-min washout gradient to 80% acetonitrile. All buffers contained 0.1% formic acid. Eluate from the inline C18 column was injected to a qExactive Plus mass spectrometer (Thermo) and top-5 MS/MS spectra collected with 1+ ions excluded from MS/MS.

FASTA libraries were generated containing only the "middle" peptides for all pairs of cleavage sites on the HdeA C18S/C66S background sequence. By definition, all these peptides began with a Cys and ended with an Arg (the nucleophile used for cleavage). Decoy libraries were generated using the De Bruyn algorithm in the Comet pipeline of the Trans Proteomic Pipeline v6.[82] All decoys were set to also begin with Cys and end with Arg, with only the sequence in-between those termini being scrambled. For additional searches using missed-cyanylation sites, separate FASTA libraries were generated for peptides spanning from the protein N-terminus to a C-terminal Arg or from Cys to the protein C-terminus, respectively. Decoys were likewise required to obey the same constraint on the termini as true-hit peptides. The C-terminal peptide library was not used further due to a very small number of missed-cyanylation peptides that could be identified in any samples. By contrast, the N-terminal missed-cyanylation library was useful due to the apparently reduced efficiency of cyanylating certain N-terminal sites, especially position 17.

The MS/MS output files were searched against these FASTA databases using Comet (TPP version 6). A required modification of the peptide N-terminal Cys residue was added to the Comet parameter file to account for the conversion of Cys to iminothiazole. Mass tolerance was set to 1.1 Da. in the Comet parameter file, to account for incorrect calling of monoisotopic peaks for larger peptides. The Comet parameter file used for the searches are included in the SI. The output.pep.xml files were manually edited to change the value of the "enzyme name" field from "CDAP" to "nonspecific," then passed to the PeptideProphet tool with default parameters except for no accurate mass binning and using only the E-value as the discriminant. Finally, this output was passed to the Quantic tool with default parameters to quantify the total ion current for top-6 MS/MS peaks for each matched peptide spectrum regardless of nominal PeptideProphet probability. The output was exported to Excel.

The false-discovery rate as a function of Comet E-value was determined for each output file empirically. Since crosslinks between sequence near-neighbors are inherently less informative for HTDS, and, moreover, short peptides may not produce a sufficient number of MS/MS ions for Quantic, data were filtered to remove any peptides where the Cys residues were closer than 12 peptide bonds apart. Data were also filtered to remove any apparent peptide spectrum matches where the identified double-Cys variant was not found in the DNA-level deep sequencing dataset (i.e., the set of 1,453 variants whose toxicity values were deemed sufficiently reliable). The remaining peptides were ranked by ascending E-value and by whether they were decoys or true hits. The false-discovery threshold was set at 10%. In other words, the top N true-hit peptides with the smallest E-values were used for further analysis, such that no more than N/10 peptide spectra from decoy hits had E-values within this range. These N peptide spectrum matches were deemed usable.

For evaluating the disulfide bonding propensities (**Figure 4**), the total ion current values for all usable peptide spectrum matches for a given variant were summed. For evaluating PSM-weighted averages of toxicity values by HIC fraction (**Figure 7g**), they were weighted by the proportion of peptide spectrum matches for that variant within that fraction. Of those, 284 (~25%) were detected in the total periplasmic library, but only 132 of the 284 were also detected in the NEM-treated library, enabling quantification of disulfide formation propensities (**Figure 4a**). The remaining variants were mostly of very low abundance (**Figure SI 13**) and could have been missed in the NEM-treated sample by chance, especially given weaker overall signal from that sample (**Figure 4b,c**). The 34 most abundant (per **Figure SI 13**) variants detected only in the total periplasmic sample but not the NEM-treated sample are plotted in gray in **Figure 4a**. Note that this experiment looked only at variants purified from periplasmic extracts; the experiment shown in **Figure 7** looked at the 339 variants that were found extracellularly. The overlap between the total periplasmic and total extracellular detected variant sets was 177 variants, so a grand total of 446 variants (~39% of theoretically possible) were detected at the protein level across all experiments in the present study. HIC fractionation of the extracellular library may have aided detection of some variants because HIC fractions with low total [protein] were concentrated prior to analysis. Violin plots in **Figure 4d** were generated using Plotly Chart Studio.

To investigate how allele toxicity relates to biophysical properties, we purified two protein library batches (expressed 2.5 months apart) from culture supernatants. Proteins from each were deep-sequenced as above. Using the empirical 10% false-discovery rate (FDR) and analyzing only middle peptides, we identified 81 variants in batch 1 and 115 in batch 2, the overlap being 56 variants. Quantification using Quantic (Comet pipeline in TPP6) showed very good batch/batch agreement (**Figure SI 15**), so the two batches were mixed and fractionated by HIC, where hydrophilic proteins elute first (i.e., at higher [ammonium sulfate]) and hydrophobic ones later. A total of 339 protein variants were detected in at least one of eight HIC fractions (**Figure 7a**): 199 from middle peptides only, 104 from N-terminal missed-cyanylation peptides only, and 36 from both. (The 10% FDR threshold was applied to each fraction separately.) The low overlap was expected because MS/MS fragmentation typically fails for long peptides (>50-60 residues), and the maximum peptide length in Comet is 63 (**SI files 1,2**); thus, e.g., WT could never be identified from N-terminal missed-cyanylation because that peptide would be 66 residues long.

*Intact protein mass spectrometry*

Electrospray-ionization isotopically resolved mass spectrometry of intact proteins was carried out as described.[83] The protein samples were analyzed on a Bruker Impact II q-TOF mass spectrometer equipped with an Agilent 1290 HPLC. The separation and desalting was performed on an Agilent PLRP-S Column (1000A, 4.6 x 50 mm, 5 µm). Mobile phase A was 0.1% formic acid in water and mobile phase B was acetonitrile with 0.1% formic acid. A constant flow rate of 0.300 ml/min was used. Ten microliters of the protein solution was injected and washed on the column for the first 2 minutes at 0%B, diverting non-retained materials to waste. The protein was then eluted using a linear gradient from 0%B to 100%B over 8 minutes. The mobile phase composition was maintained at 100%B for 1 minutes and then returned to 0%B over 0.1 minute. The column was re-equilibrated to 0%B for the next 5.9 minutes. A plug of sodium formate was introduced at the end of the run, to perform internal m/z calibration to obtain accurate m/z values. The data were analyzed using Bruker Compass DataAnalysis™ software (Version 4.3, Build 110.102.1532, 64 bit). The charge state distribution for the protein produced by electrospray ionization was deconvoluted to neutral charge state using DataAnalysis implementation of the Maximum Entropy algorithm. Predicted isotope patterns were calculated at the resolving power of 50,000 and compared with isotopically resolved neutral mass spectra calculated using Maximum Entropy from the experimental charge state distribution.

*Intrinsic Trp fluorescence*

Fluorescence spectra were measured at room temperature in a Cary Eclipse fluorimeter (Varian/Agilent) with excitation at 290 nm (10 nm slit) and emission scanned from 300 to 450 nm (5 nm slit) and 700 V PMT voltage. Protein samples were prepared at 9-10 µM (identical within each set of 4 samples) in 10 mM pH 7 sodium phosphates buffer with 150 mM NaCl, unless otherwise indicated. Four samples were read in parallel using a multi-cuvette holder; no-protein blanks in the same cuvettes were subtracted. Five spectra per sample were averaged.

*Bis-ANS fluorescence*

To the same protein samples that were used for measuring intrinsic fluorescence, bisANS was added to 10 µM final concentration. Excitation wavelength was 400 nm (10 nm slit); emission was scanned 450-600 nm (5 nm slits) at room temperature. PMT voltage was kept at 700 V. For experiments with reducing agent, TCEP was added to 5 mM final concentration. Sample dilutions were no more than 5% from addition of the compounds.

*Circular dichroism*

Ellipticity was measured on a Jasco J-1500 instrument, at room temperature, in 10 mM pH 7 sodium phosphates buffer, in a 1 mm path length cuvette with 16-20 µM [protein] and 50 nm/min scanning rate, with 5 scans averaged per sample. Curves were smoothed by taking the moving average with a 3-nm sliding window.

*Size-exclusion chromatography / Multiangle light scattering (SEC/MALS)*

An SRT SEC-150 column (Sepax) was pre-equilibrated with 10 mM pH 7 sodium phosphates with 150 mM NaCl on an Infinity 2 1260 HPLC system (Agilent) with an inline degasser. The absolute refractive index value for the buffer was determined empirically. The inline UV detector, the Dawn Heleos II multiangle light scattering detector (Wyatt), and the OptilabTrEX RI detector (Wyatt) were aligned using a sample of bovine serum albumin. HdeA samples (20 µM, 100 µL) were injected and run at 0.5 ml/min for 45 min., with MALS detection every 0.5 s. Signals from the highest two and lowest two angle detectors were removed to reduce noise. Molecular weights were determined automatically using ASTRA 7 software, with "Normal" despiking and manually adjusted baselines where required.

*Atomistic Monte-Carlo simulations*

We used our previously established MCPU software, which uses a knowledge-based, statistical potential for atomistic modeling of protein unfolding intermediates,[37,39,50] the MCPU version used in this study is available at https://github.com/proteins247/dbfold. To mimic the effect of a disulfide linkage between two cysteines, we use a strong flat-bottomed harmonic potential as a restraint between the CB atoms of a pair of cysteine residues. For the restraint, the region of zero potential consists of CB-CB distances between 2.9 and 4.6 Å.[84] The restraint force constant was set to 200 $Å^{-2}$. An NMR structure of E. coli HdeA (PDB ID: 5WYO) was used as the initial structural model. A 13-residue linker consisting of alternating G and S residues was built to connect the C-terminus of the A chain to the N-terminus of the B chain in the homodimer structure. This linker is needed due to a limitation in the simulation software. For each double-Cys variant, Modeller (version 9.24)[85] was used to mutate target residues to cysteine. Original cysteine residues 18 and 66 were mutated to serine as necessary. NAMD (version 2.13),[86] with the CHARMM 22 forcefield,[87] was used to minimize the initial structure. Then, each minimized cysteine variant was simulated in a multiplexed temperature replica exchange MCPU simulation with flat-bottom harmonic restraints active between the two intramolecular cysteine pairs. Each simulation used 21 temperatures (between 0.350 and 0.850, inclusive, in steps of 0.025) with four replicas at each temperature. Simulations were run for 1.275 billion MC steps. A knowledge-based moveset was enabled for the first 300 million steps. The replica exchange interval was set to 10,000 steps, and simulation structures were saved every 1,000,000 steps.

Simulation data from the last 275 million steps were used for analysis. Simulation samples from between temperatures of 0.375 and 0.625 were analyzed. Simulation structures were clustered based on intermolecular (inter-subunit) contact maps. For each simulation sample, a 2D Boolean matrix corresponding to contacts between residues 11 to 89 of one subunit and residues 11 to 89 of the other subunit was constructed, with a contact defined as a CA-CA distance of less than 10 Å. The flattened matrix was then used as a feature vector for clustering. The Jaccard distance metric was used as the measure of dissimilarity, and clustering was performed using the DBSCAN algorithm,[51] as implemented in scikit-learn.[88] For DBSCAN, epsilon was set to 0.57 and minimum points to 14. Clustering was performed on structures from all simulated variants. For computational tractability, the number of samples from simulations was reduced 10-fold (i.e. structures from every 10 million steps such that each replica provides 28 structures). Simulation observables such as distances, secondary structure, and radius of gyration were measured using the Python package MDTraj.[89] Expectation values for observables at each temperature were calculated using the pymbar package.[54]

For more in-depth simulations of the free energy landscapes of select variants, we used monomeric starting structures (subunit A of PDB ID 5WYO), since the mutants were shown experimentally to be monomeric, and applied umbrella biasing as well as replica exchange for enhanced sampling. The mutations were generated and monomeric variants minimized using MOE.[90] From each minimized structure, we simulated the variant at 20 MCPU temperatures between 0.250 and 0.725 for 600 million MC steps (including knowledge-based moves for the first 50 million steps). The simulation structures along with their energy values were saved every 500,000 steps.

Data from the last 400 million steps were used for analysis. The pymbar package was used to calculate the free energy landscape as a function of the fraction of native contacts (Q) and RMSD. RMSD and native contacts fraction (using the definition from [91]) were computed using the Python package MDTraj.[89] Surface hydrophobicity for each simulation frame was defined as the sum of Miyazawa–Jernigen hydrophobicity values of each residue[92] weighted by the solvent accessible surface area (SASA) of that residue normalized to total SASA of the protein. By thermal averaging the surface hydrophobicity of each snapshot using the free energy difference calculated with the pymbar, the expectation value of the hydrophobicity was measured.

**Acknowledgments**
E. S. is grateful to Dr. Bharat V. Adkar and Dr. Joao V. Rodrigues for mentoring in the generation and cloning of DNA variant libraries and for insightful discussions.

This work was supported by the National Institutes of Health grants F32GM126651 and K99GM141459 to E. S. and R35GM139571 to E. I. S.

## Supplementary Figures

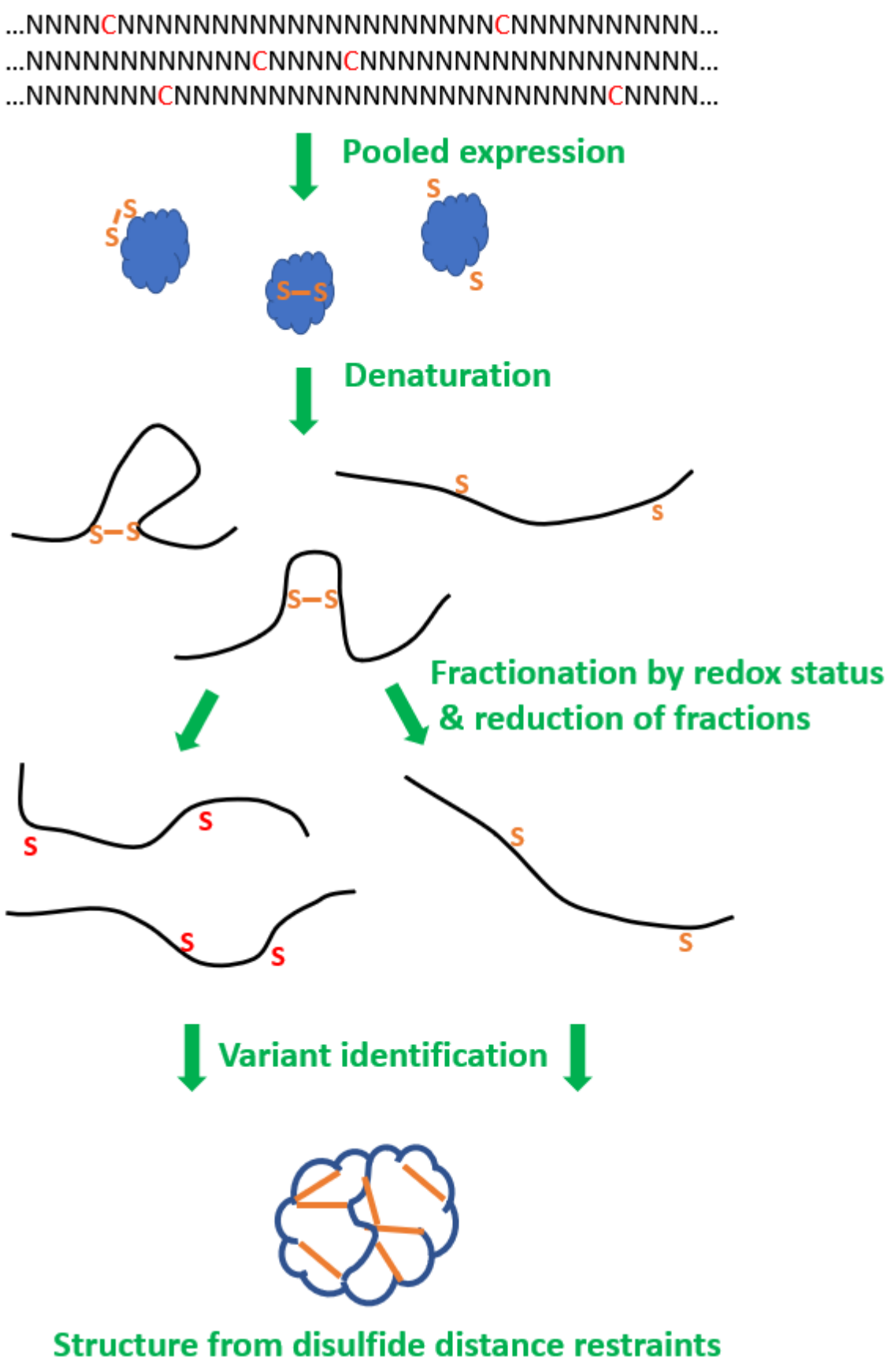

**Figure SI 1: A hypothetical illustration of protein structure determination by high-throughput disulfide scanning (HTDS).** The essence of the technique is to reduce the 3D problem of protein structure mapping to the 1D problem of protein sequencing. First, a pooled double-Cys scanning library is expressed, and as the proteins fold, pairs of Cys are brought sufficiently close together in the tertiary structure to form disulfides, while other pairs are separated and remain reduced. This effectively encodes the 3D structure of the protein in a 2D matrix of disulfide formation propensity as a function of the two sequence positions. The tertiary structure can then be denatured without loss of structural information, provided the disulfides remain intact. The library is then fractionated by redox status. This can be accomplished either by thiol affinity chromatography or – as in our study – by simply alkylating all free thiols at this step. This effectively encodes the 2D matrix of disulfide propensities in the 1D sequences of the protein variants. Disulfides can then be reduced without loss of structural information, provided the polypeptide backbones remain intact. The disulfide-forming variants are then identified by protein sequencing. In the current version of our method, this means cyanylation of the Cys thiols followed by aminolysis of the polypeptide backbone immediately upstream of the cyano-Cys sites. MS/MS then identifies and quantifies each double-Cys variant by measuring the middle fragment from this Cys-targeted backbone cleavage. Finally, a computational method is required to back-calculate the original 3D structure from the 2D matrix of disulfide bonding propensities.

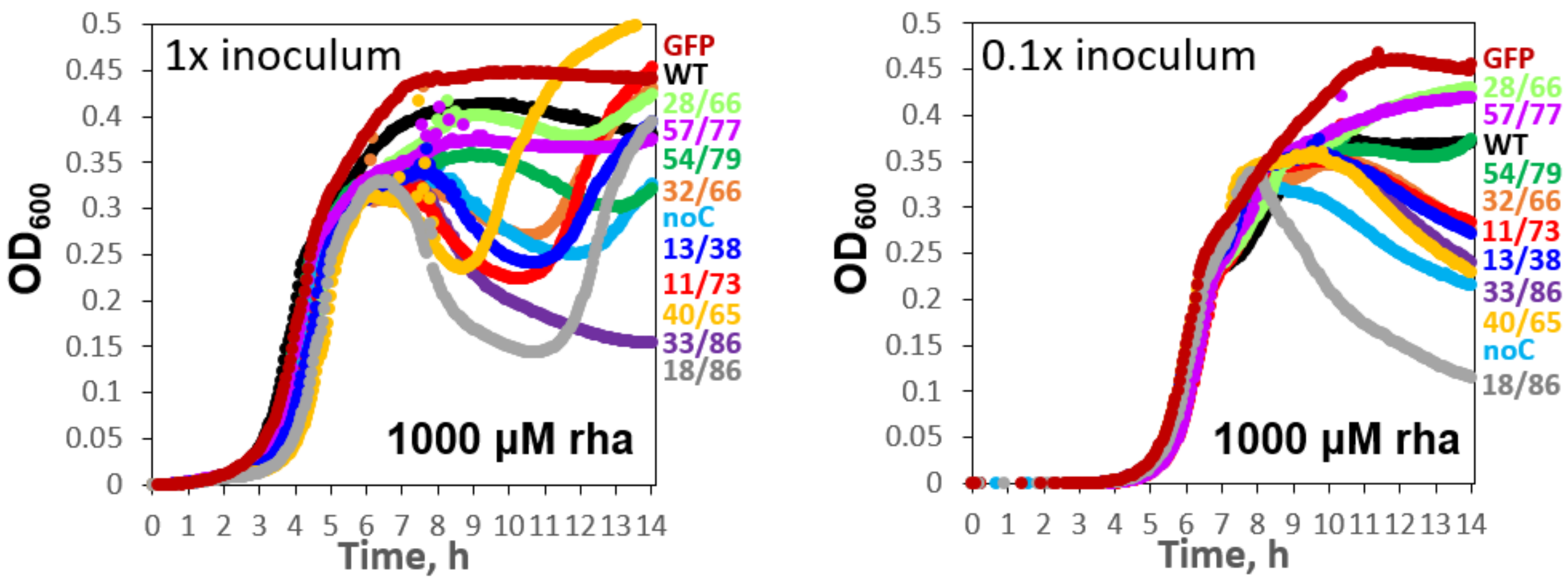

**Figure SI 2: Effect of inoculum size on the growth curve dynamics of HdeA variant-expressing Δ*hdea* BW25113 *E. coli*.** *Left*: the same plot as **Figure 2a**, shown for reference. *Right*: the equivalent experiment with 1/10 the inoculum size.

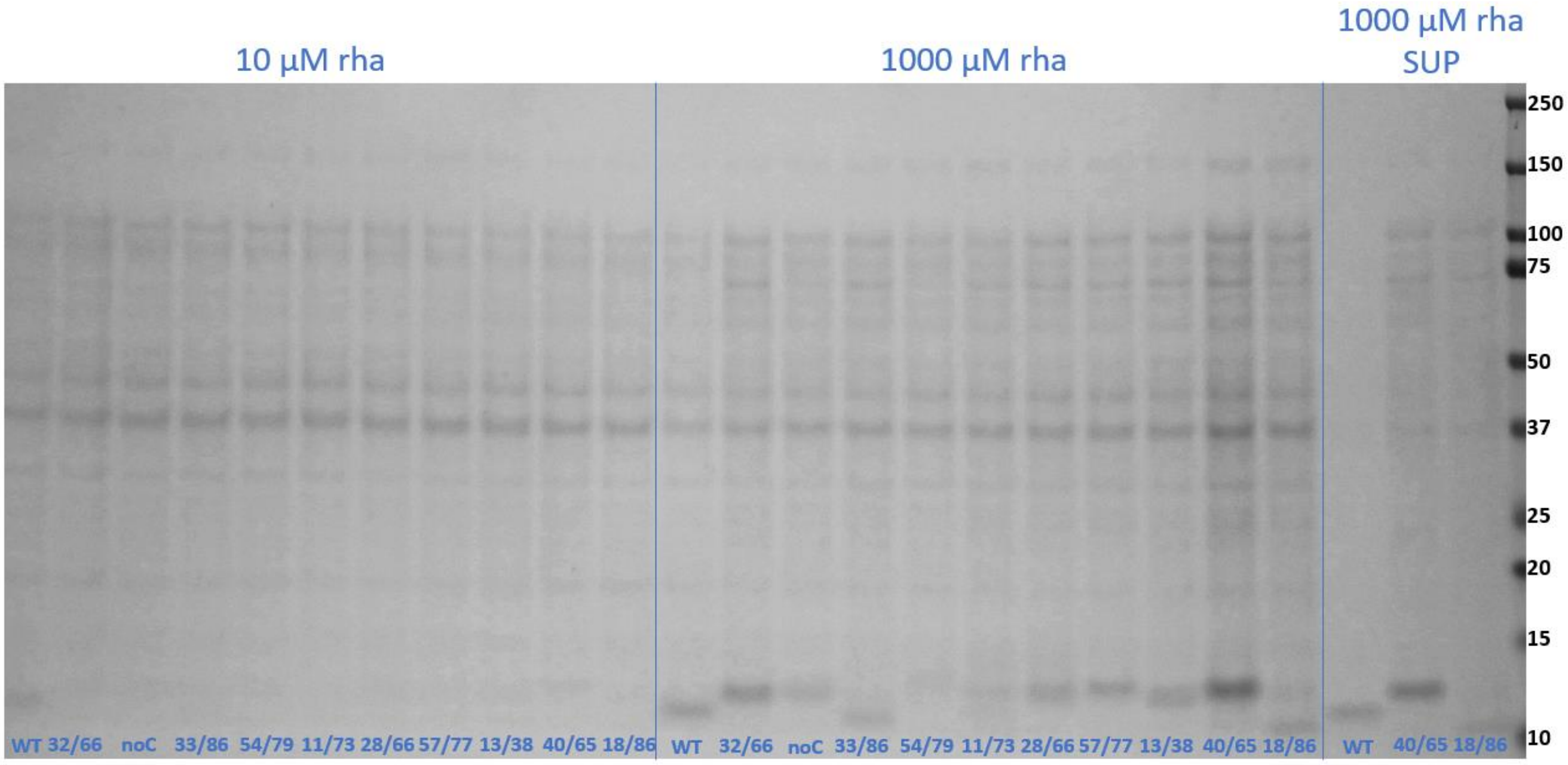

**Figure SI 3: Non-reducing SDS-PAGE of total end-point cultures grown and induced as in Figure 2b, but without using MOPS buffer in the LB medium during initial (pre-induction) overnight culture.** Expression levels were more variable under these conditions; at 10 µM rhamnose, only the WT was visibly expressed, but at 1000 µM rhamnose, all variants had visible HdeA bands consistent with their expected migration patterns.

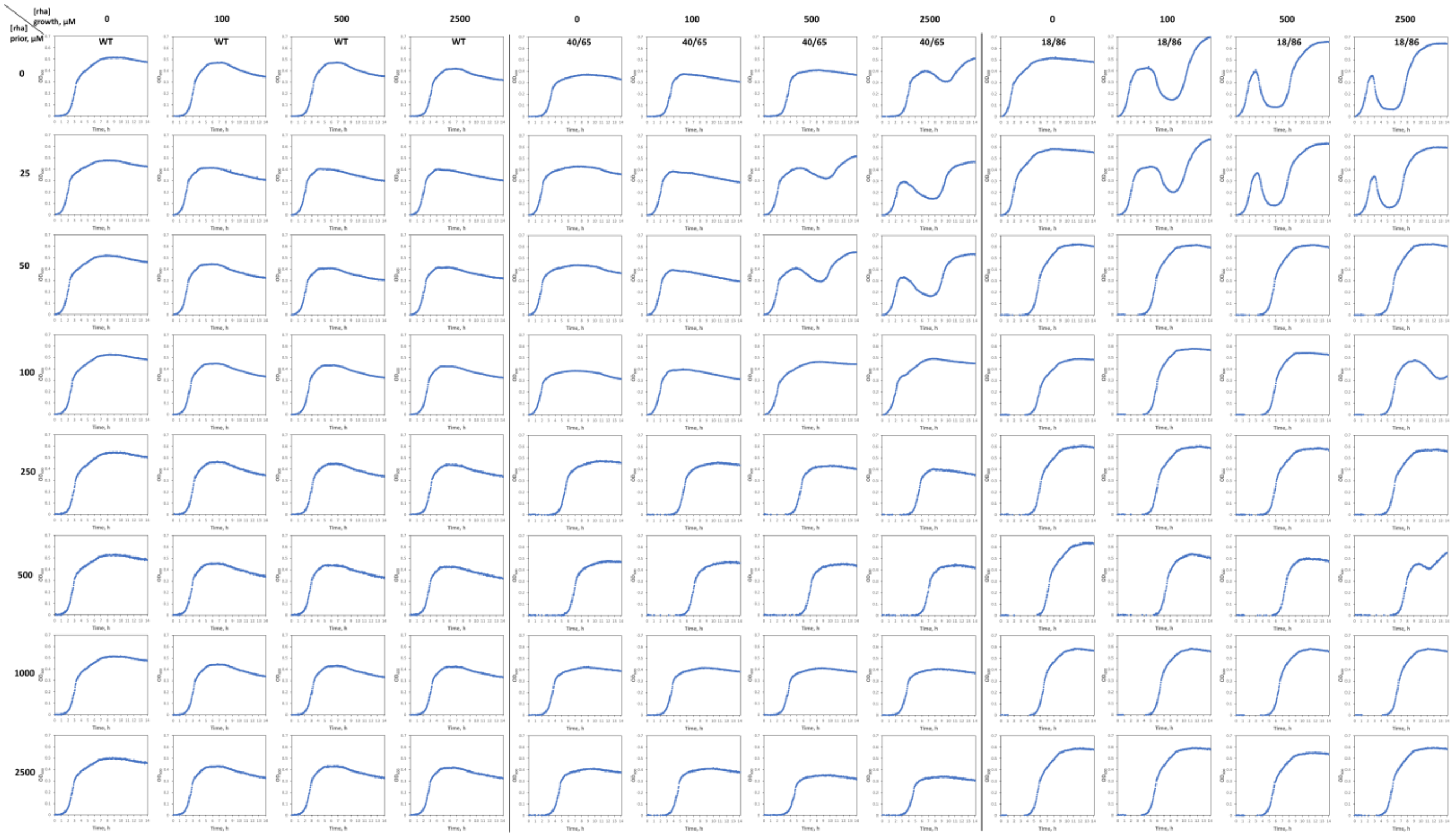

**Figure SI 4: *E. coli* cultures show persistent adaptation after recovery from cytotoxic HdeA variant expression.** Three variants were compared: WT HdeA (left four columns), variant 40/65 (middle four columns), and variant 18/86 (right four columns). Numbers from top to bottom indicate [inducer] during initial culture, from 0 to 2500 µM. Numbers across the top indicate [inducer] during a subsequent culture inoculated from the initial culture: 0, 100, 500, or 2500 µM for each variant. As expected, when the initial culture was done without inducer (*top row*), the growth curves of the two cytotoxic variants (but not of the WT) showed a strong die-off dependent on induction level, followed by recovery. However, when initial cultures contained inducer (*subsequent rows*), and the second culture was inoculated from the recovered initial cultures, then the die-off phenotype disappeared. However, these recovered cultures had visibly longer lag phases than the naïve cultures, suggesting an epigenetic memory that alters growth dynamics.

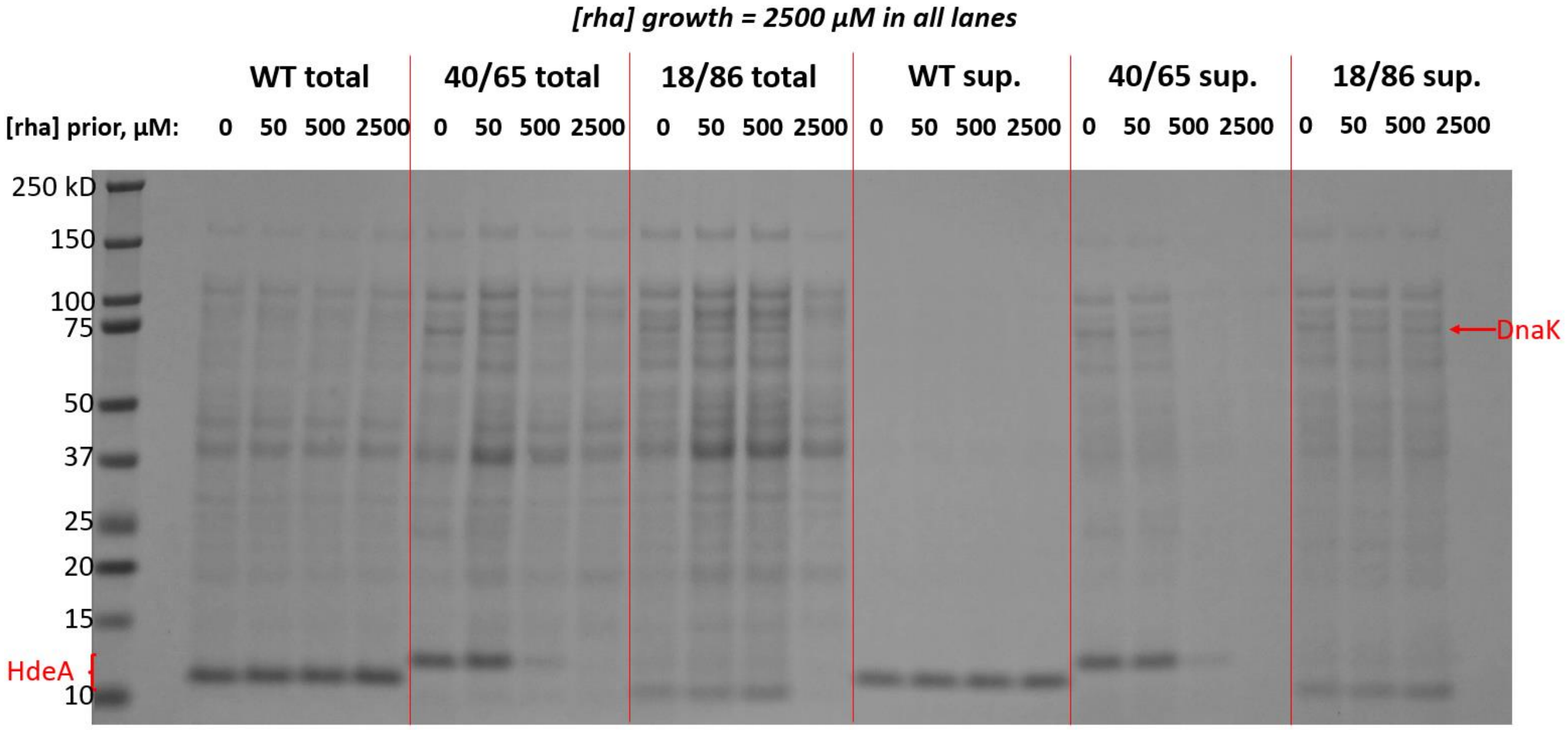

**Figure SI 5: Recovery from cytotoxicity caused by HdeA misfolding entails loss of HdeA expression.** Non-reducing SDS-PAGE of total cell culture and cell culture supernatants from the endpoints of the respective growth curves shown in columns 4, 8, and 12 of **Figure SI 3** (i.e., secondary cultures grown in the presence of 2500 µM rhamnose). Inducer concentrations during primary culture were as indicated: 0, 50, 500, or 2500 µM. WT HdeA cultures continued to express the protein ("WT total" lanes), but 40/65 and 18/86 cultures lost protein expression if the primary cultures contained high [rhamnose] ("40/65 total" and "18/86 total" lanes). Moreover, the overexpression of endogenous DnaK (*red arrow*) likewise diminished, as did cytolysis (the "sup." lanes).

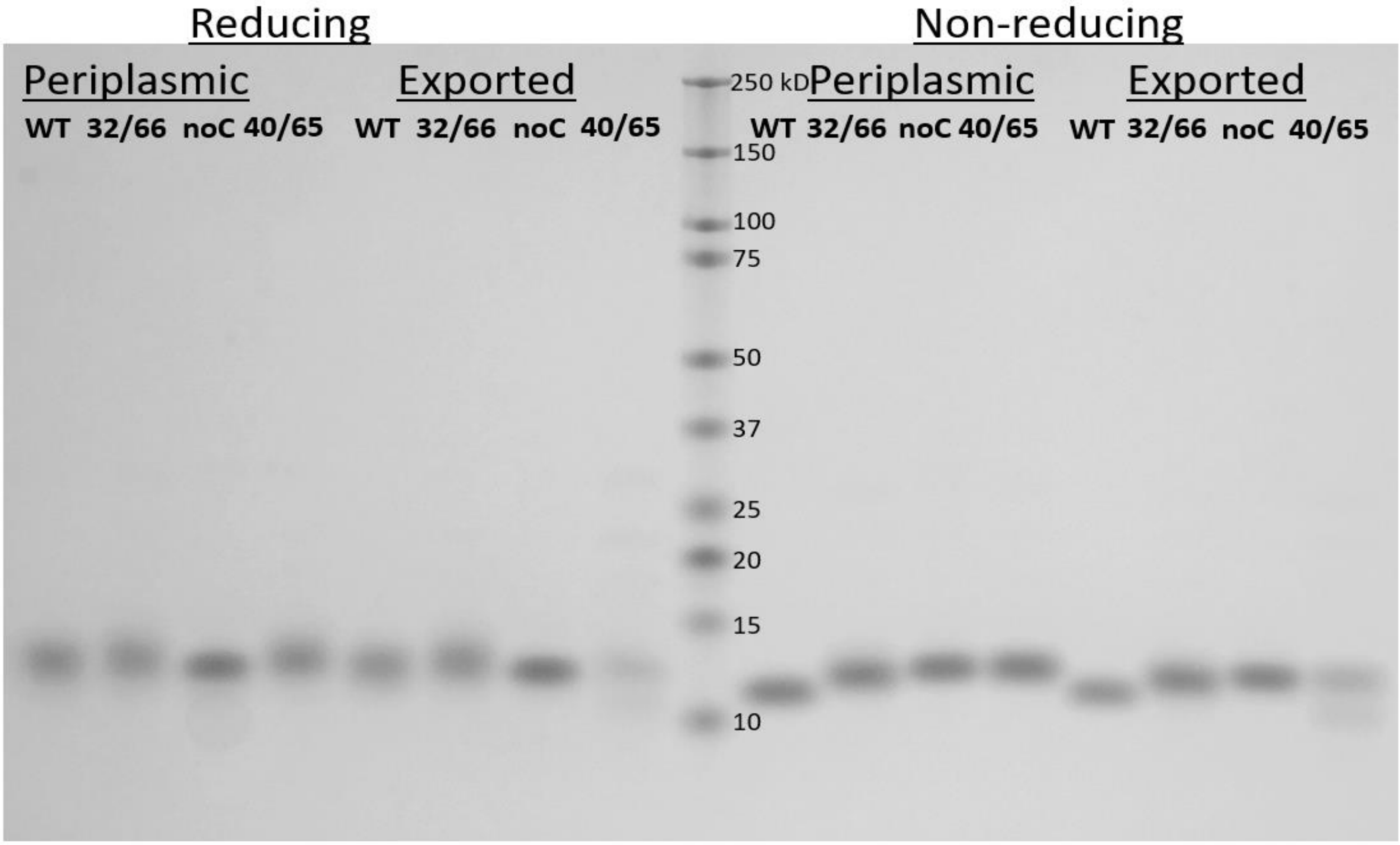

**Figure SI 6: Effect of disulfide bonds on HdeA variants' migration in SDS-PAGE.** Proteins were purified both from periplasmic extracts ("Periplasmic") and from the culture supernatants ("Exported"). Samples on the left side of the gel were treated with reducing agent (5 mM TCEP) during the denaturation step; samples on the right side were denatured without reduction.

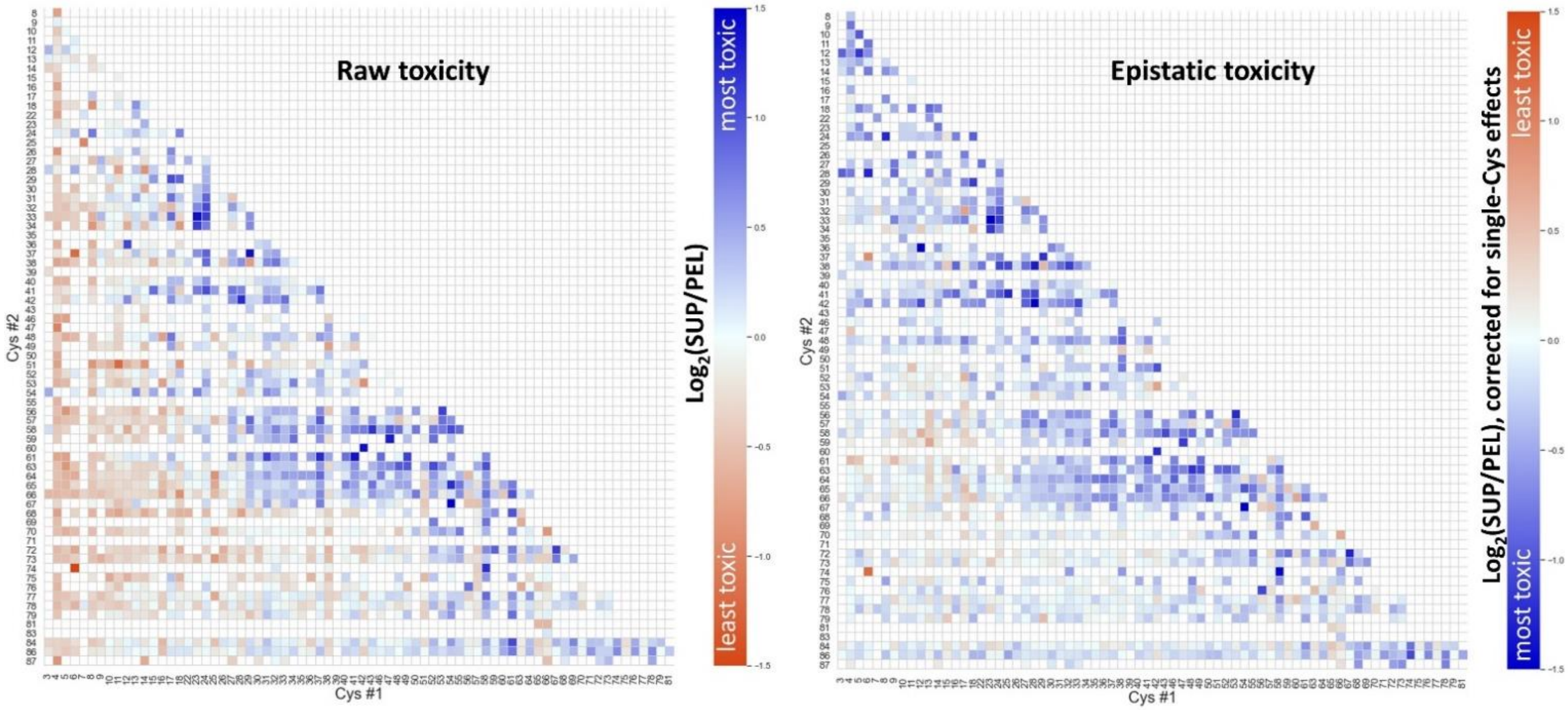

**Figure SI 7: Landscapes of allele toxicity of double-Cys HdeA variants.** *Left*: The raw average values of allele toxicity, as expressed by $\log_2$ ratios of the given allele's abundance in the culture supernatant ("SUP") vs. the cell pellet ("PEL"), showed that many Cys mutations in the disordered N-terminal region (approx. residues 1-14) resulted in rescue of toxicity compared to the noC variant (defined as 0 in this figure), but a clear cluster of highly toxic variants (approx. residues 29-49 x 58-66) was still evident. *Right*: When these raw values were corrected by dividing them by the product of toxicity effects of single-Cys variants at the corresponding positions, the signal from the N-terminal region largely disappeared, revealing a toxicity map quite similar to **Figure 3b**.

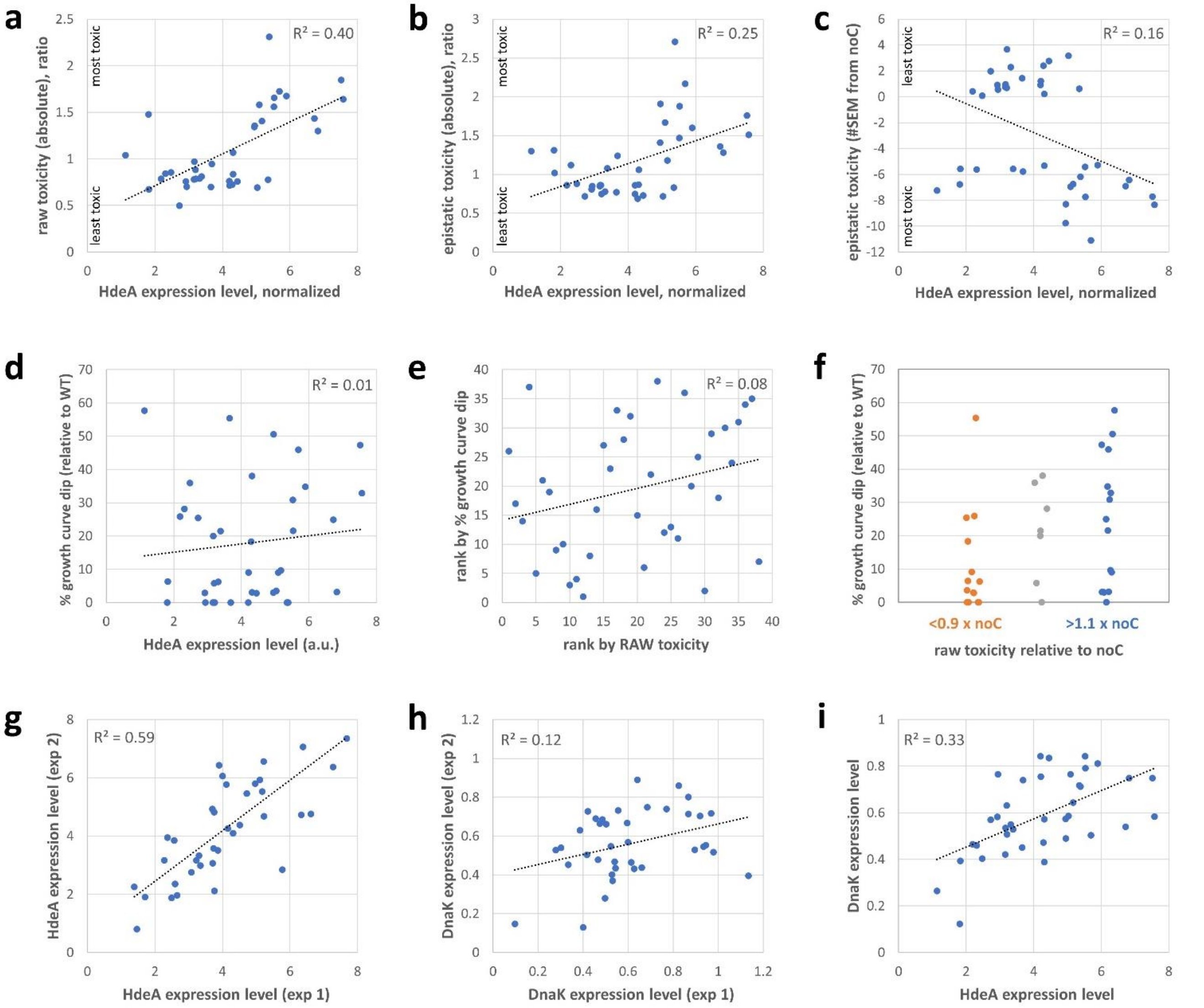

**Figure SI 8: Low-throughput characterization of expression level effects in 38 randomly chosen double-Cys HdeA variants.** Two independent expression experiments were conducted, with growth curves measured by $OD_{600}$ in 96-well format in sealed plates for 14 h and expression levels assayed by non-reducing SDS-PAGE from endpoint total culture samples from the same plates. HdeA and DnaK expression levels were internally normalized to the most intense band (37 kD) within each lane. Correlation between HdeA variant expression level and its raw allele toxicity **(a)** was much stronger than with epistatic toxicity **(b)**, especially if expressed as #SEM from noC **(c)**, the metric we use in the main text. Notably, there was no correlation between the magnitude of the dip in the growth curve and the HdeA expression level **(d)** and only a very weak correlation with the raw toxicity rank **(e)**. However, when variants were grouped by raw toxicity, the difference between group 1 (<90% of noC's toxicity) and group 3 (>110% of noC's toxicity) appeared significant by the one-tailed Mann-Whitney U test ($p = 0.025$). HdeA variant expression levels were reproducible between the two experiments **(f)**, but less so the DnaK expression levels **(g)**, which we attribute to the much fainter DnaK bands and therefore much higher noise in quantitating them. For all other panels, the expression levels from the two independent experiments were averaged to reduce noise. The correlation between HdeA and DnaK expression levels **(h)** should be considered as probably strong, given the high uncertainty in [DnaK] estimates **(g)**.

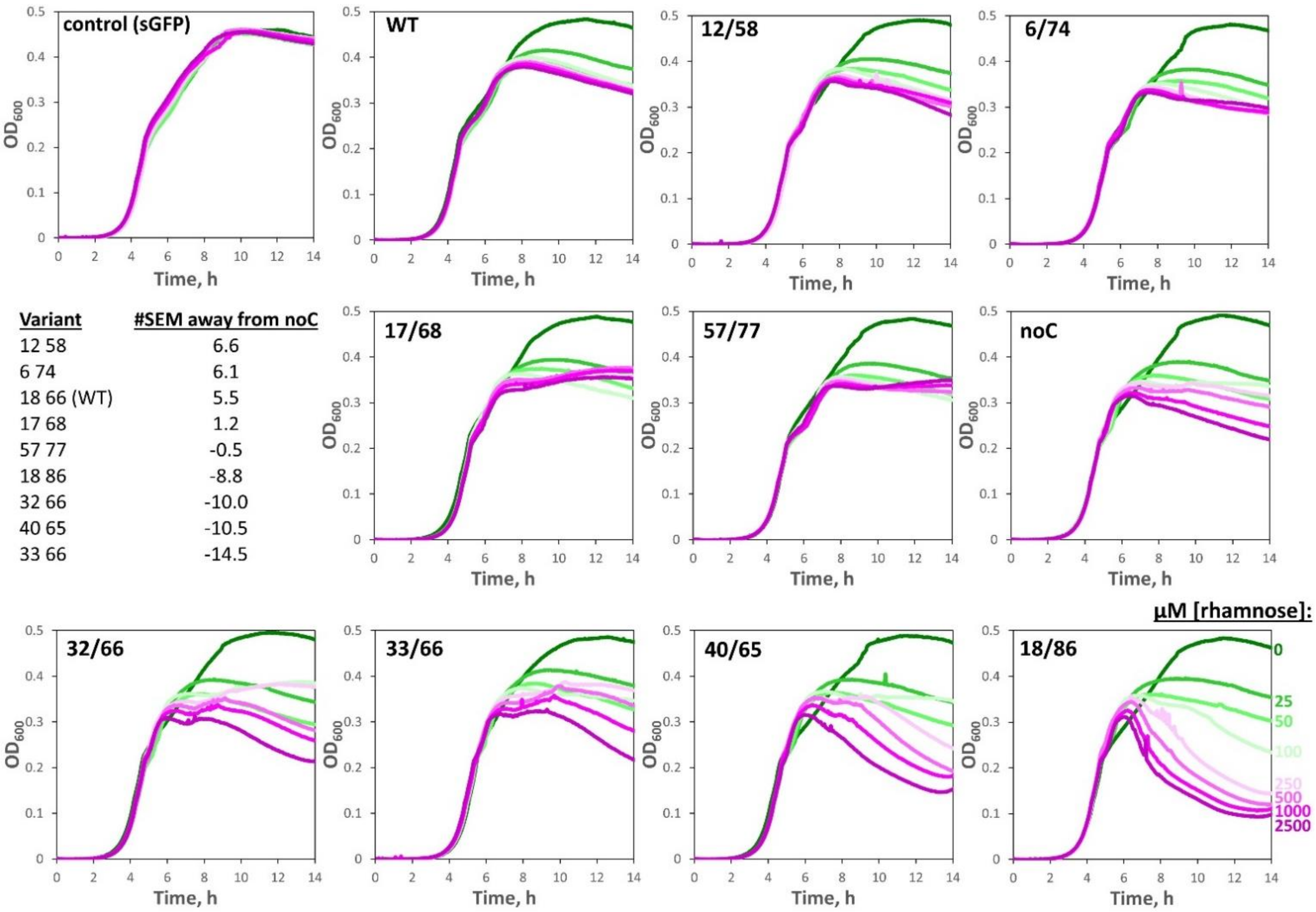

| Variant | #SEM away from noC |
|---|---|
| 12 58 | 6.6 |
| 6 74 | 6.1 |
| 18 66 (WT) | 5.5 |
| 17 68 | 1.2 |
| 57 77 | -0.5 |
| 18 86 | -8.8 |
| 32 66 | -10.0 |
| 40 65 | -10.5 |
| 33 66 | -14.5 |

**Figure SI 9: Dose-dependent growth curve shapes of select HdeA variants.** Growth curves of polyclonal transformant cultures in LB broth with antibiotics were monitored as in **Figure SI 2** with the indicated concentrations of inducer (rhamnose) present in the culture medium. While all HdeA variants showed some reduction of stationary-phase optical density when the proteins were expressed relative to no expression, the 12/58 and 6/74 variants were the only ones with WT-like traces, validating the results of the high-throughput assay in **Figure 3c**. The next-least-toxic variants, 17/68 and 57/77, showed a cusp upward in the stationary-phase growth curve at high induction. We did not investigate the origin of this effect, but it could be attributable, e.g., to cell lengthening induced by a mild heat shock – which would be consistent with pronounced cusps upward in the more toxic 32/66 and 40/65 variants at lower induction levels.

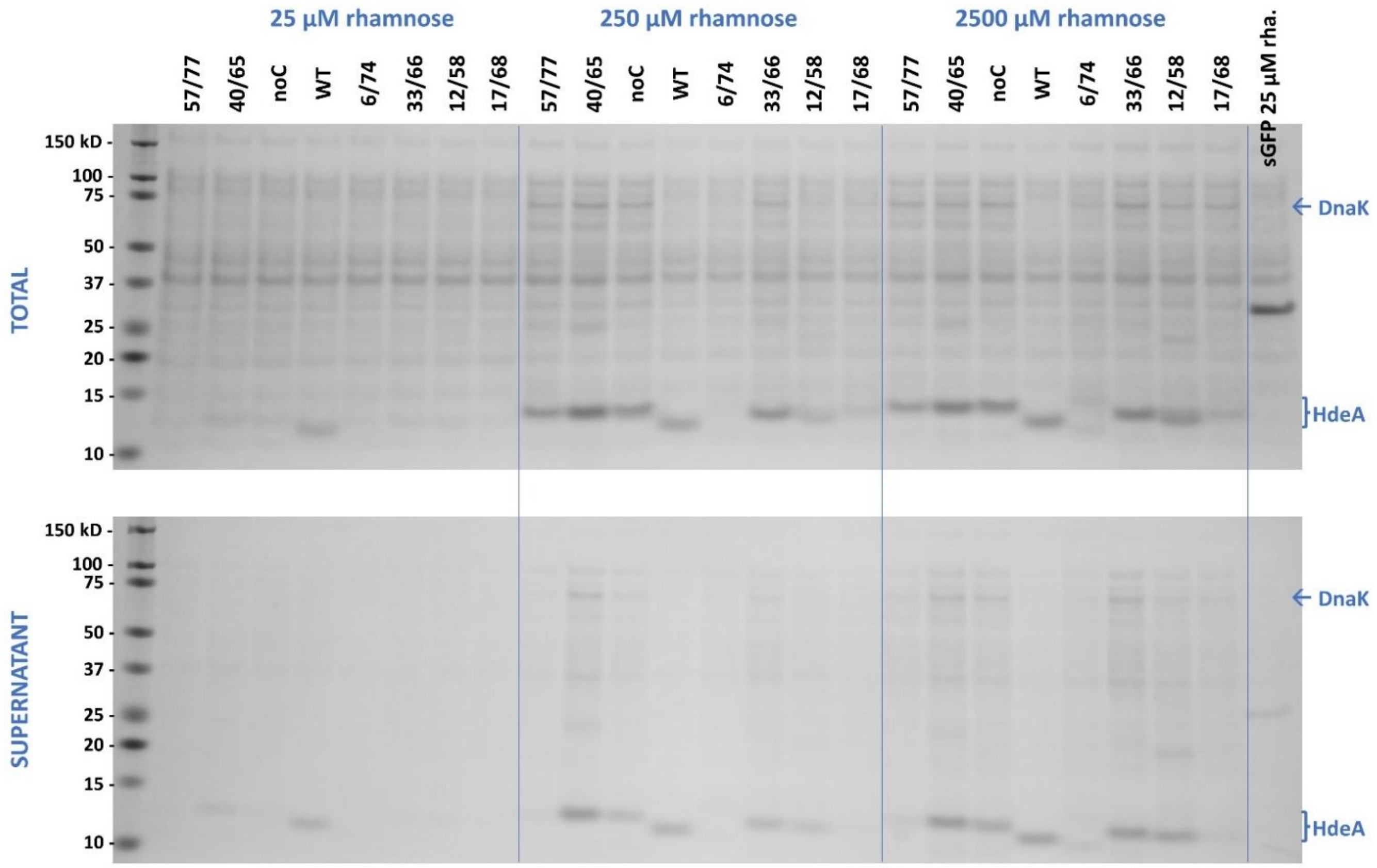

**Figure SI 10: Some non-toxic non-native double-Cys variants have exceptionally low levels of expression.** Non-reducing SDS-PAGE of total (*top gel*) and supernatant (*bottom gel*) samples from the endpoints of the growth curves in Figure SI 9, at 25, 250, and 2500 µM rhamnose, as indicated. The total culture samples showed greatly reduced expression of 6/74 and 17/68 relative to the WT, noC, or the toxic variants 40/65 and 33/66. Variant 12/58 also had a much reduced expression level at low or medium [inducer], but did accumulate to a significant level at the highest [inducer]. At the highest [inducer], however, both DnaK overexpression and cell lysis became evident even for that variant. (Note that culture spindown was not completely effective in this case, as evidenced by a trace of sGFP in the control lane on the bottom gel, as well as minor traces of cellular proteins in the supernatant of WT cultures; however, the much greater lysogenicity of the toxic variants is still visually apparent.) At the highest level of induction, even DnaK overexpression was visible in every HdeA variant except the WT. One unexpected observation was the near-total lack of export of the 57/77 variant from the overexpressing cells, despite significant accumulation comparable to the WT and the toxic mutants. We hypothesize that this variant may be retained in the cytoplasm of the cells, perhaps even in inclusion bodies, though other possibilities exist.

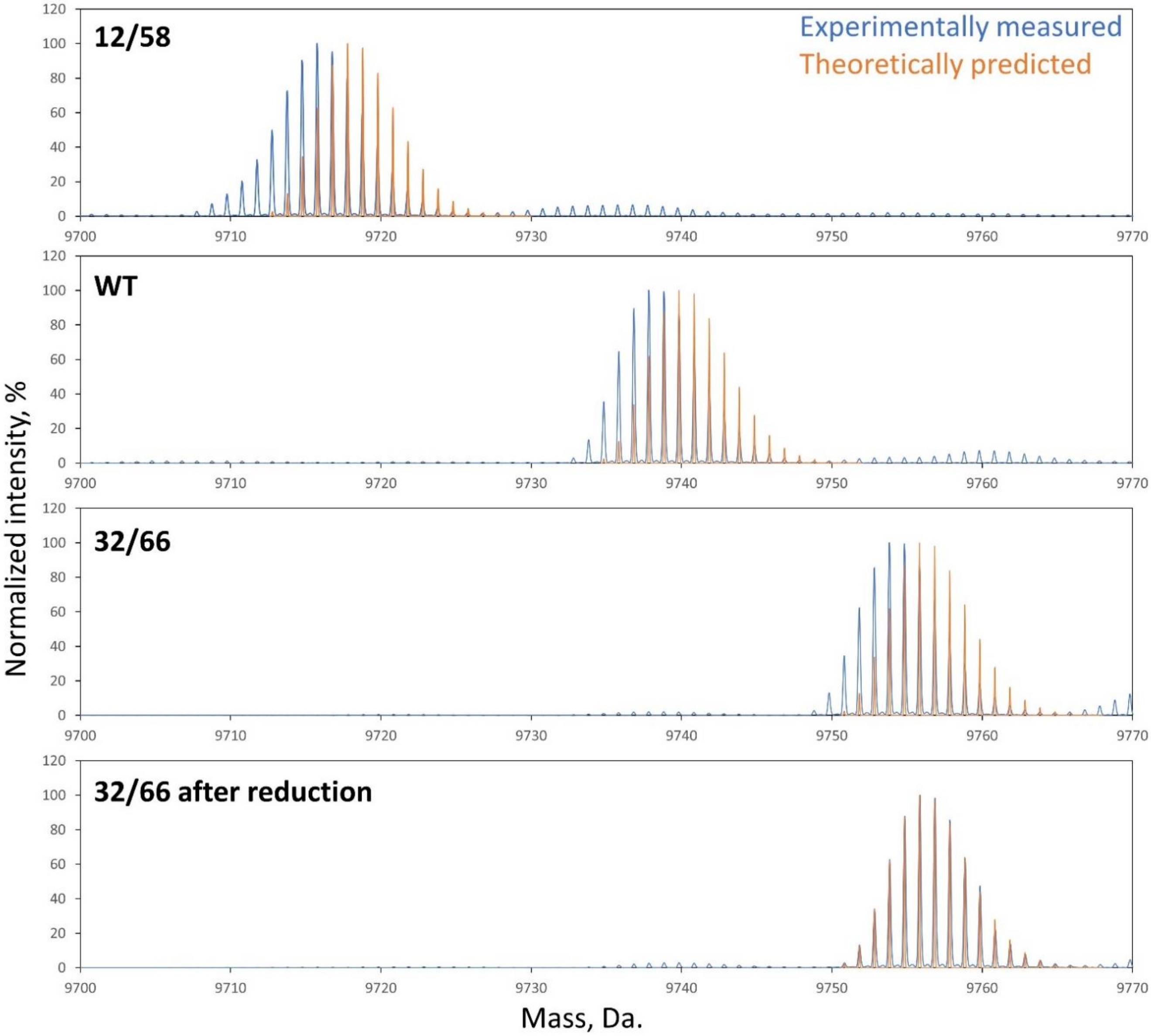

**Figure SI 11: Isotopically resolved electrospray mass spectrometry of intact HdeA variants confirms disulfide bond formation.** WT HdeA, as well as the 32/66 and 12/58 variants, were purified entirely in the disulfide-bonded state, as evidenced by the clear -2 Da. shift in the isotopic distribution compared to predicted values for the fully reduced proteins. The loss of two protons indicates oxidation of two -SH groups to form a disulfide. As a control (*bottom graph*), we carried out the same experiment with the same 32/66 sample after treating it with a reducing agent, which eliminated the -2 Da. shift. Minor peaks at +22 Da. were observed in all samples, attributable to a rare Na+ adduct in place of a proton.

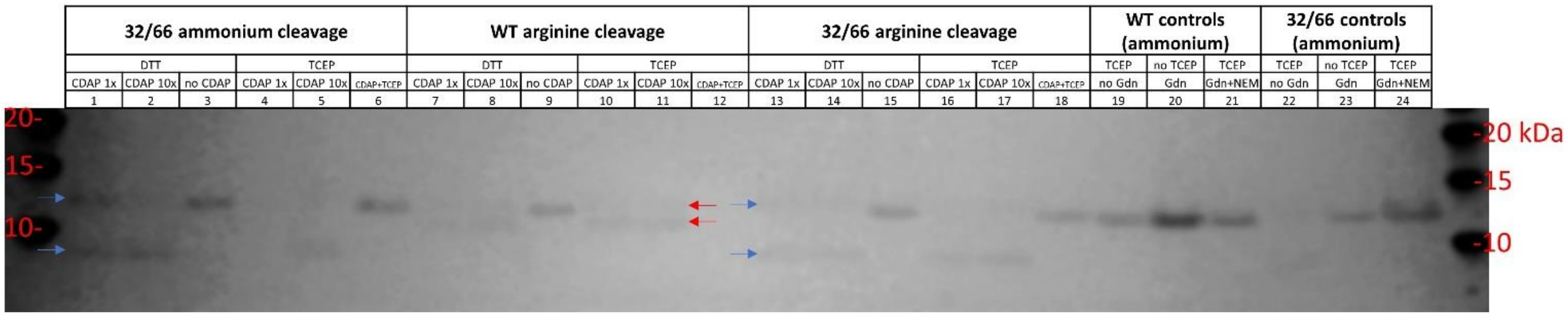

**Figure SI 12: Targeted cleavage of cyano-Cys containing HdeA by aminolysis.** Two variants (WT and 32/66) were used, since they yield fragments of distinct length upon Cys-specific cleavage. Prior to cyanylation, samples were reduced by either dithiothreitol (DTT) or tricarboxyethyl phosphine (TCEP), as indicated, with subsequent removal of the reducing agent by ultrafiltration. Either ammonium (3 M final concentration) or arginine (3 M final concentration) was used as the nucleophile for cleavage, as labeled. All ammonium-cleaved samples also contained 1 M (final concentration) of guanidinium chloride to further aid denaturation. All cleavage was at pH 9. Since samples containing high levels of guanidinium or arginine are not compatible with SDS-PAGE, all samples were desalted using C18 tips, with a wash step at 5% acetonitrile and elution in 50% acetonitrile, with 0.1% formic acid in all cases. The variable efficiency of elution and relatively low binding capacity of the C18 tips resulted in low and variable band intensities on the resulting gel; therefore, band intensities cannot be compared across lanes but should only be compared within each lane. Blue and red *arrows* indicate the intact (upper) and longest fragment (lower) bands for the 32/66 and WT (18/66) proteins, respectively. Several important results are evident by comparison of certain sets of lanes:

*Lanes 1 and 2* show that higher [CDAP] at the cyanylation step led to higher yield at the aminolysis step, suggesting that the limiting factor for aminolysis was the efficiency of cyanylation. “1x” = 1 mM.

*Lanes 3, 9, and 15* confirm that the pH 9 incubation with the nucleophiles did not by itself cause any detectable cleavage, meaning that cleavage was specific to cyano-Cys as expected.

*Lanes 2, 5, 11, 14, and 17* indicate that, at least at the higher [CDAP], overall reaction yield was very high. (Note that, since the cleavage fragment is shorter, it is expected to bind fewer Coomassie molecules per mole than the full-length protein, yet the intensities of the fragment bands in those lanes were higher than of the respective full-length bands.)

*Lanes 6, 12, and 18* indicate that presence of the reducing agent TCEP (added back after the ultrafiltration step) inhibited the cyanylation reaction, even though TCEP does not contain thiols (unlike DTT). Thus, removal of the reducing agent was important for reaction efficiency and is therefore preferable to the “one-pot” approach when high-efficiency cleavage is a priority.

*Lanes 19 and 22* show that the WT was not efficiently cleaved in the absence of initial denaturation (“no Gdn”), attributable to inaccessibility of the native disulfide bond in the folded protein to the reductant; by contrast, the 32/66 variant, with its predominantly molten structure, was still cleaved.

*Lanes 20 and 23* confirm that in the absence of initial reduction, no subsequent cleavage occurs; thus, only free thiols were modified by the cyanylating agent.

*Lanes 21 and 24* show that when thiol-blocking reagent (N-ethyl maleimide, NEM) was present during the initial reduction step with TCEP, no subsequent cleavage occurred, either; thus, Cys residues that are not oxidized but covalently modified by thiol-blocker were likewise not susceptible to cyanylation and cleavage.

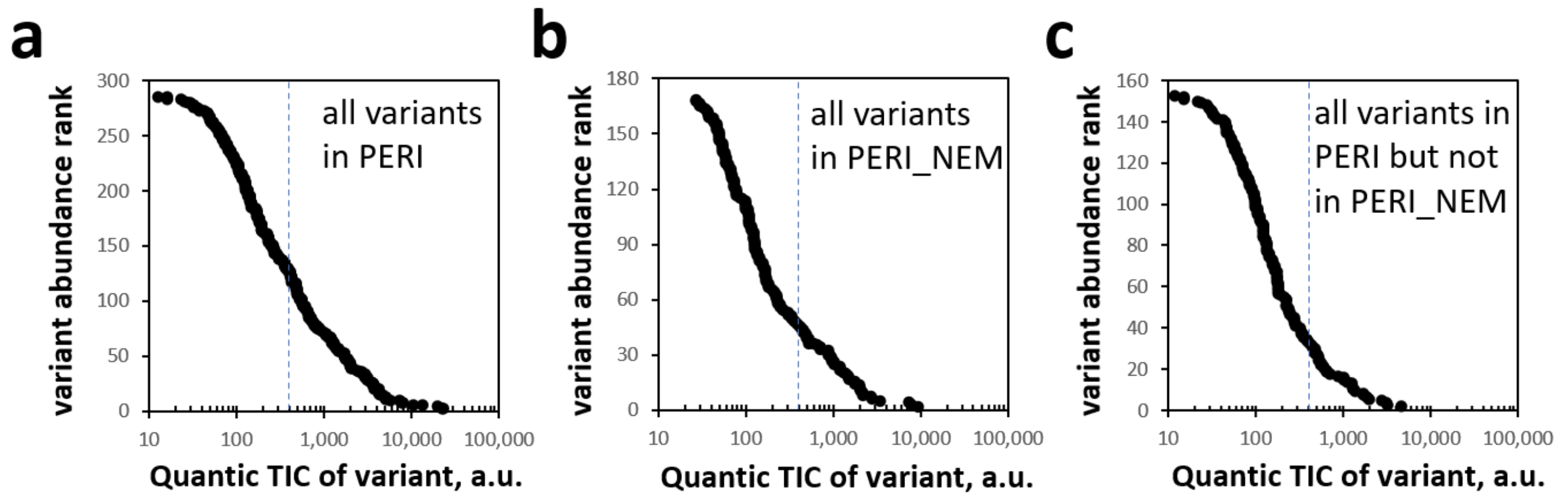

**Figure SI 13: Total ion current distributions for variants detected in Figure 4.** Peptide abundances are estimated by total ion current (TIC) using the Quantic tool of TPP6 Comet pipeline, based on intensities of top-6 MS/MS fragment ions. **(a)** Log-linear plot of all variants in PERI ranked by abundance revealed a sigmoidal distribution of abundance. **(b)** Same as **a** but for PERI_NEM. **(c)** Same as **a** but for variants found in PERI and not in PERI_NEM, showing that most were low-abundance variants (weak total ion currents). A blue dashed line at 400 a.u. is shown in all three plots for reference.

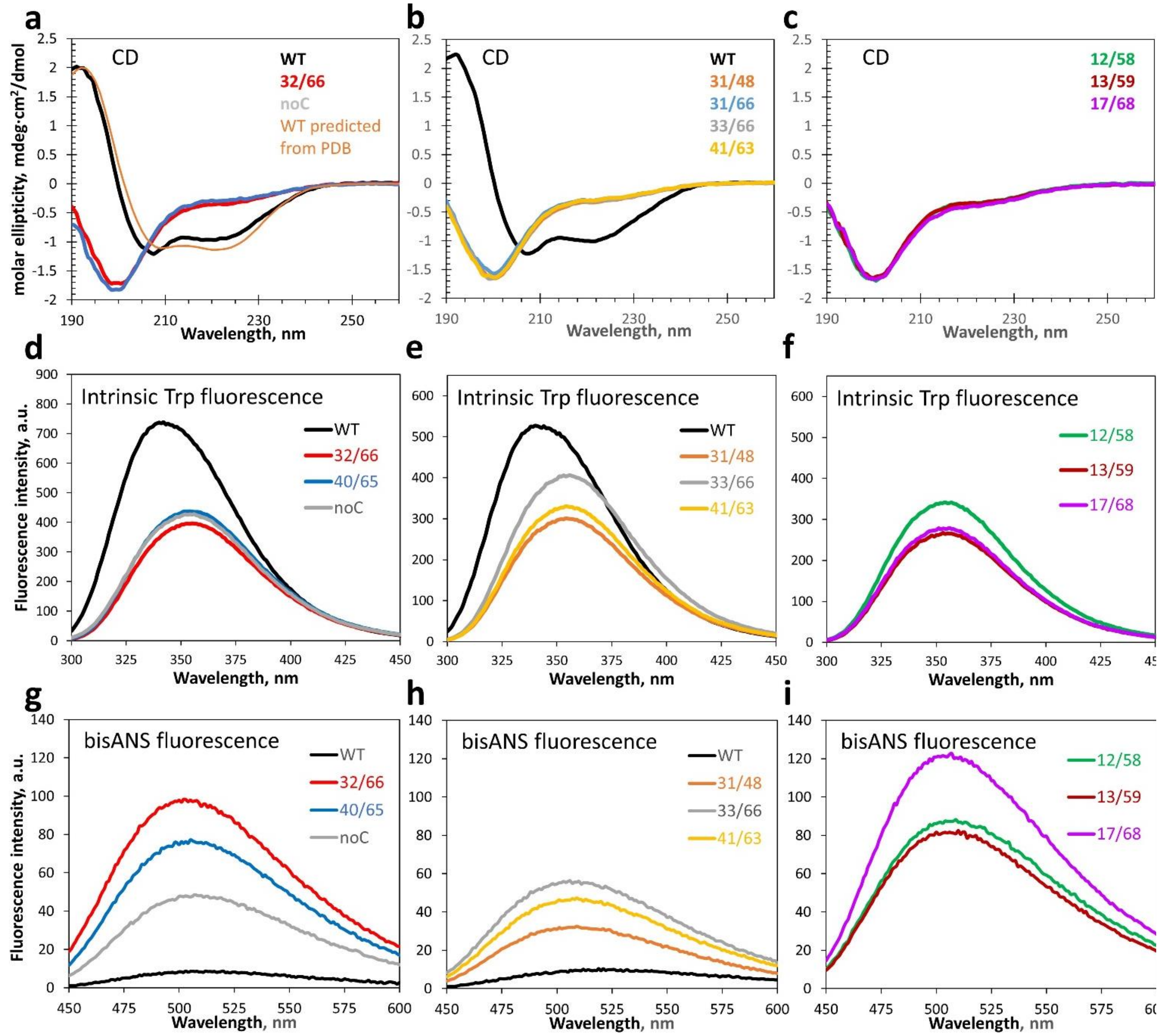

**Figure SI 14: Double-Cys variants do not rescue the native folded structure.** Panels **a**, **d**, and **g** are identical to **Figure 6c,a,d** and show the circular dichroism, intrinsic fluorescence, and bisANS fluorescence spectra for the WT, noC, 32/66, and 40/65 variants. Panels **b**, **e**, and **h** show the same data for the 31/48, 33/66, and 41/63 variants, as well as a separate sample of WT as an internal control. Panel **b** also includes the CD spectrum of the 31/66 variant. Panels **c**, **f**, and **i** show the same data for the 12/58, 13/59, and 17/68 variants.

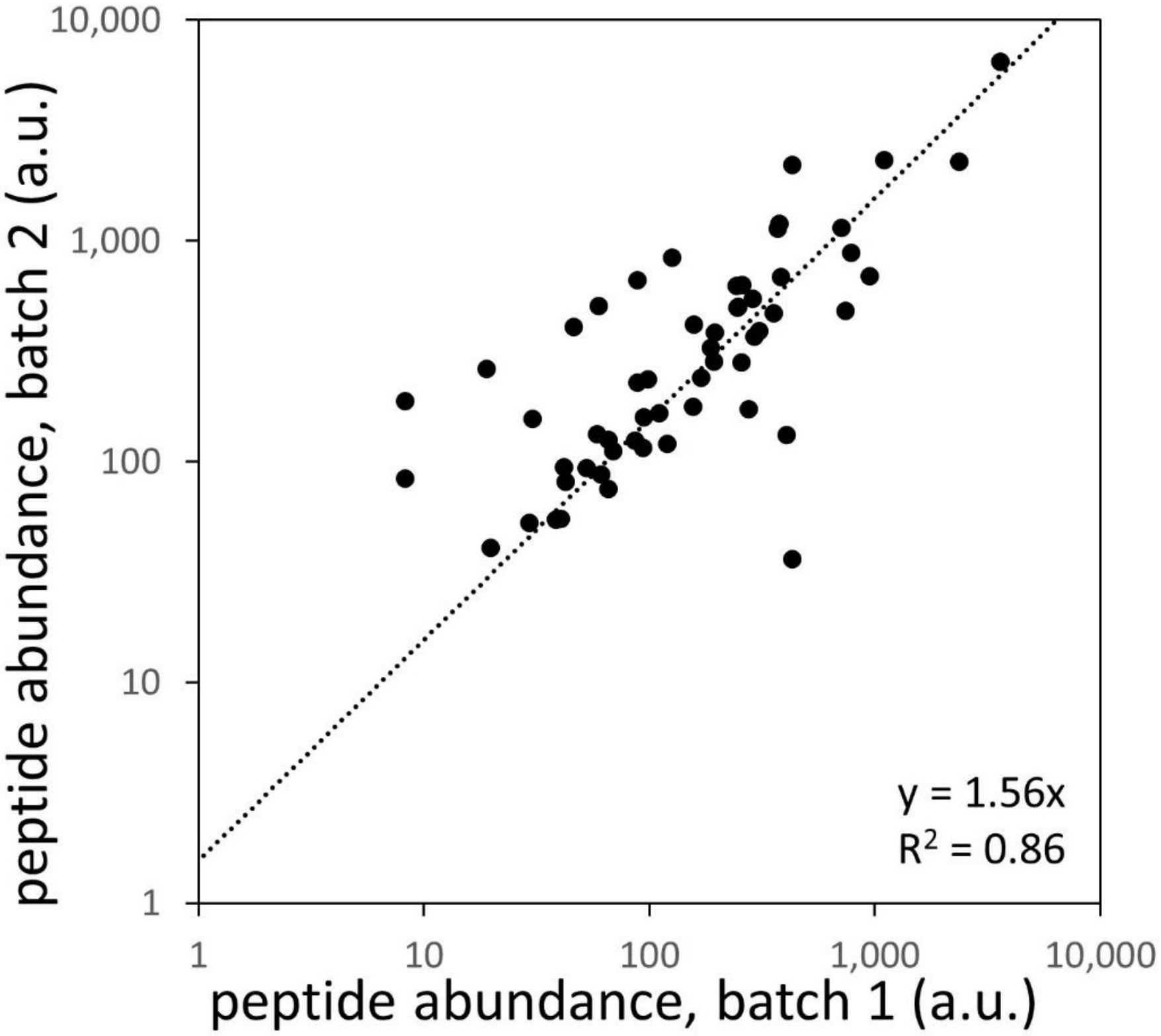

**Figure SI 15: Label-free peptide quantification by LC/MS/MS is reproducible.** Peptide abundances quantified by the Quantic tool of TPP6 Comet are plotted for the 56 variants identified in two separate batches (full biological replicates) of the multi-Cys scanning library. The true ratio of total protein concentrations injected onto LC/MS/MS for the two batches was 1.40, while the line of best fit returned a ratio of 1.56.

**Table SI 1: primer sequences**

fixed reverse primer: 5’-GAACTCAGAAGTGAAACG-3’

short mutagenic forward primers (mutated codon in **bold font**), 5'-3'
(a silent mutation in template is in red)

| | |
|---|---|
| noCssHdeAC3 | GCAGCCGAT**TGC**CAAAAAGCAGCT |
| noCssHdeAC4 | GCCGATGCG**TGC**AAAGCAGCTGAT |
| noCssHdeAC5 | GATGCGCAA**TGC**GCAGCTGATAAC |
| noCssHdeAC6 | GATGCGCAAAAA**TGC**GCTGATAAC |
| noCssHdeAC7 | CAAAAAGCA**TGC**GATAACAAAAAACCG |
| noCssHdeAC9 | GCAGCTGAT**TGC**AAAAAACCGGTC |
| noCssHdeAC10 | GCTGATAAC**TGC**AAACCGGTCAAC |
| noCssHdeAC11 | GATAACAAA**TGC**CCGGTCAACTCC |
| noCssHdeAC13 | CAAAAAACCG**TGC**AACTCCTGG |
| noCssHdeAC14 | CAAAAAACCGGTC**TGC**TCCTGGACC |
| noCssHdeAC15 | CCGGTCAAC**TGC**TGGACCAGT |
| noCssHdeAC17 | GTCAACTCCTGG**TGC**AGTGAAGATTTC |
| noCssHdeAC18 | TCCTGGACC**TGC**GAAGATTTCCTG |
| noCssHdeAC23 | GATTTCCTG**TGC**GTGGACGAATCC |
| noCssHdeAC24 | GATTTCCTGGCT**TGC**GACGAATCC |
| noCssHdeAC27 | GTGGACGAA**TGC**TTCCAGCCAACT |
| noCssHdeAC29 | GAATCCTTC**TGC**CCAACTGCA |
| noCssHdeAC31 | CTTCCAGCCA**TGC**GCAGTTGGT |
| noCssHdeAC32 | CAGCCAACT**TGC**GTTGGTTTTGCT |
| noCssHdeAC33 | CCAACTGCA**TGC**GGTTTTGCTGAAGCG |
| noCssHdeAC36 | GCAGTTGGTTTT**TGC**GAAGCGCTG |
| noCssHdeAC38 | GGTTTTGCTGAA**TGC**CTGAACAAC |
| noCssHdeAC40 | GAAGCGCTG**TGC**AACAAAGATAAACCA |
| noCssHdeAC41 | GAAGCGCTGAAC**TGC**AAAGATAAACCA |
| noCssHdeAC42 | GCGCTGAACAAC**TGC**GATAAACCA |
| noCssHdeAC48 | CCAGAAGAT**TGC**GTTTTAGATGTTCAG |
| noCssHdeAC49 | CCAGAAGATGCG**TGC**TTAGATGTTCAG |
| noCssHdeAC52 | GTTTTAGAT**TGC**CAGGGTATTGCAACC |
| noCssHdeAC53 | GTTTTAGATGTT**TGC**GGTATTGCAACC |
| noCssHdeAC56 | CAGGGTATT**TGC**ACCGTAACC |
| noCssHdeAC57 | GGTATTGCA**TGC**GTAACCCCA |
| noCssHdeAC58 | ATTGCAACC**TGC**ACCCCAGCTATC |
| noCssHdeAC59 | ATTGCAACCGTA**TGC**CCAGCTATC |
| noCssHdeAC61 | GTAACCCCA**TGC**ATCGTTCAGGCT |
| noCssHdeAC63 | CCAGCTATC**TGC**CAGGCTAGTACT |
| noCssHdeAC64 | GCTATCGTT**TGC**GCTAGTACTCAG |
| noCssHdeAC65 | GCTATCGTTCAG**TGC**AGTACTCAG |
| noCssHdeAC66 | GTTCAGGCT**TGC**ACTCAGGATAAACAAGCC |
| noCssHdeAC67 | CAGGCTAGT**TGC**CAGGATAAACAAGCC |
| noCssHdeAC68 | CAGGCTAGTACT**TGC**GATAAACAAGCC |
| noCssHdeAC70 | GCTAGTACTCAGGAT**TGC**CAAGCCAAC |
| noCssHdeAC71 | CAGGATAAA**TGC**GCCAACTTTAAAGAT |
| noCssHdeAC72 | CAGGATAAACAA**TGC**AACTTTAAAGAT |
| noCssHdeAC73 | GATAAACAAGCC**TGC**TTTAAAGATAAAGTT |
| noCssHdeAC75 | GCCAACTTT**TGC**GATAAAGTTAAAGGC |
| noCssHdeAC77 | GCCAACTTTAAAGAT**TGC**GTTAAAGGC |
| noCssHdeAC78 | CTTTAAAGATAAA**TGC**AAAGGCGAATGG |
| noCssHdeAC79 | CTTTAAAGATAAAGTT**TGC**GGCGAATGG |
| noCssHdeAC84 | GGCGAATGGGAC**TGC**ATTAAGAAAGAT |
| noCssHdeAC86 | GAATGGGACAAAATT**TGC**AAAGATATG |
| noCssHdeAC87 | GACAAAATTAAG**TGC**GATATGTAACACCATCAC |

**Table SI 2: HIC and toxicity values for 39 most abundant variants**

| variant | HIC peak, raw | HIC peak, corrected | #SEM from noC |
|---|---|---|---|
| 18/66 (WT) | 2.1 | 2.1 | 5.5 |
| 70/84 | 2.3 | 2.3 | -1.0 |
| 37/61 | 4.7 | 4.7 | -2.2 |
| 24/57 | 5.7 | 5.7 | -2.3 |
| 31/58 | 4.9 | 4.9 | -1.6 |
| 24/49 | 6.0 | 6.0 | -1.5 |
| 34/76 | 5.5 | 5.5 | -0.5 |
| 31/57 | 5.5 | 5.5 | -5.7 |
| 33/66 | 6.8 | 6.8 | -14.5 |
| 29/57 | 6.3 | 6.3 | -7.7 |
| 24/38 | 6.0 | 6.0 | -8.4 |
| 18/32 | 6.2 | 6.2 | -1.6 |
| 29/66 | 7.0 | 7.5 | -6.9 |
| 32/66 | 6.0 | 6.5 | -10.0 |
| 34/66 | 5.4 | 7.5 | -9.8 |
| 42/66 | 2.2 | 2.2 | -6.0 |
| 47/66 | 4.7 | 4.7 | -5.7 |
| 17/29 | 6.9 | 6.9 | -3.2 |
| 17/30 | 6.9 | 6.9 | -0.7 |
| 17/31 | 6.5 | 6.5 | -2.2 |
| 17/32 | 6.1 | 6.1 | 2.1 |
| 17/33 | 6.1 | 6.1 | -5.5 |

Color coding indicates which panel of Figure 7 of the main text shows the HIC trace.

**SI File 1: Comet (TPP6) parameters for middle peptides search**

```
# comet_version 2021.01 rev. 0
# Comet MS/MS search engine parameters file.
```

```
# Everything following the '#' symbol is treated as a comment.

database_name = /some/path/db.fasta
decoy_search = 0                    # 0=no (default), 1=concatenated search, 2=separate search
peff_format = 0                     # 0=no (normal fasta, default), 1=PEFF PSI-MOD, 2=PEFF Unimod
peff_obo = C:/TPP/conf/PSI-MOD.obo                   # path to PSI Mod or Unimod OBO file

num_threads = 0                     # 0=poll CPU to set num threads; else specify num threads directly (max 128)

#
# masses
#
peptide_mass_tolerance = 1.1
peptide_mass_units = 0              # 0=amu, 1=mmu, 2=ppm
mass_type_parent = 1                # 0=average masses, 1=monoisotopic masses
mass_type_fragment = 1              # 0=average masses, 1=monoisotopic masses
precursor_tolerance_type = 1        # 0=MH+ (default), 1=precursor m/z; only valid for amu/mmu tolerances
isotope_error = 3                   # 0=off, 1=0/1 (C13 error), 2=0/1/2, 3=0/1/2/3, 4=-8/-4/0/4/8 (for +4/+8 labeling)

#
# search enzyme
#
search_enzyme_number = 11           # choose from list at end of this params file
search_enzyme2_number = 0           # second enzyme; set to 0 if no second enzyme
num_enzyme_termini = 2              # 1 (semi-digested), 2 (fully digested, default), 8 C-term unspecific , 9 N-term
unspecific
allowed_missed_cleavage = 0         # maximum value is 5; for enzyme search

#
# Up to 9 variable modifications are supported
# format:  <mass> <residues> <0=variable/else binary> <max_mods_per_peptide> <term_distance> <n/c-term>
<required> <neutral_loss>
#     e.g. 79.966331 STY 0 3 -1 0 0 97.976896
#
variable_mod01 = 15.9949 M 0 3 -1 0 0 0.0
variable_mod02 = 26.02 C 1 1 0 2 1 0.0
variable_mod03 = 0.984 DE 0 1 -1 0 0 0.0
max_variable_mods_in_peptide = 3
require_variable_mod = 0

#
# fragment ions
#
# ion trap ms/ms:  1.0005 tolerance, 0.4 offset (mono masses), theoretical_fragment_ions = 1
# high res ms/ms:    0.02 tolerance, 0.0 offset (mono masses), theoretical_fragment_ions = 0,
spectrum_batch_size = 15000
#
fragment_bin_tol = 1.0005           # binning to use on fragment ions
fragment_bin_offset = 0.4           # offset position to start the binning (0.0 to 1.0)
theoretical_fragment_ions = 1       # 0=use flanking peaks, 1=M peak only
```

use_A_ions = 0
use_B_ions = 1
use_C_ions = 0
use_X_ions = 0
use_Y_ions = 1
use_Z_ions = 0
use_Z1_ions = 0
use_NL_ions = 1 # 0=no, 1=yes to consider NH3/H2O neutral loss peaks

#
# output
#
output_sqtfile = 0 # 0=no, 1=yes write sqt file
output_txtfile = 0 # 0=no, 1=yes write tab-delimited txt file
output_pepxmlfile = 1 # 0=no, 1=yes write pepXML file
output_mzidentmlfile = 0 # 0=no, 1=yes write mzIdentML file
output_percolatorfile = 0 # 0=no, 1=yes write Percolator pin file
print_expect_score = 1 # 0=no, 1=yes to replace Sp with expect in out & sqt
num_output_lines = 5 # num peptide results to show

sample_enzyme_number = 11 # Sample enzyme which is possibly different than the one applied to the search.
# Used to calculate NTT & NMC in pepXML output (default=1 for trypsin).

#
# mzXML parameters
#
scan_range = 0 0 # start and end scan range to search; either entry can be set independently
precursor_charge = 0 0 # precursor charge range to analyze; does not override any existing charge; 0 as 1st entry ignores parameter
override_charge = 0 # 0=no, 1=override precursor charge states, 2=ignore precursor charges outside precursor_charge range, 3=see online
ms_level = 2 # MS level to analyze, valid are levels 2 (default) or 3
activation_method = ALL # activation method; used if activation method set; allowed ALL, CID, ECD, ETD, ETD+SA, PQD, HCD, IRMPD, SID

#
# misc parameters
#
digest_mass_range = 500.0 10000.0 # MH+ peptide mass range to analyze
peptide_length_range = 5 63 # minimum and maximum peptide length to analyze (default 1 63; max length 63)
num_results = 100 # number of search hits to store internally
max_duplicate_proteins = -1 # maximum number of additional duplicate protein names to report for each peptide ID; -1 reports all duplicates
max_fragment_charge = 5 # set maximum fragment charge state to analyze (allowed max 5)
max_precursor_charge = 9 # set maximum precursor charge state to analyze (allowed max 9)
nucleotide_reading_frame = 0 # 0=proteinDB, 1-6, 7=forward three, 8=reverse three, 9=all six
clip_nterm_methionine = 0 # 0=leave sequences as-is; 1=also consider sequence w/o N-term methionine

spectrum_batch_size = 15000 # max. # of spectra to search at a time; 0 to search the entire scan range in one loop
decoy_prefix = DECOY_ # decoy entries are denoted by this string which is pre-pended to each protein accession
equal_I_and_L = 1 # 0=treat I and L as different; 1=treat I and L as same
output_suffix = # add a suffix to output base names i.e. suffix "-C" generates base-C.pep.xml from base.mzXML input
mass_offsets = # one or more mass offsets to search (values substracted from deconvoluted precursor mass)
precursor_NL_ions = # one or more precursor neutral loss masses, will be added to xcorr analysis

#
# spectral processing
#
minimum_peaks = 10 # required minimum number of peaks in spectrum to search (default 10)
minimum_intensity = 0 # minimum intensity value to read in
remove_precursor_peak = 0 # 0=no, 1=yes, 2=all charge reduced precursor peaks (for ETD), 3=phosphate neutral loss peaks
remove_precursor_tolerance = 1.5 # +- Da tolerance for precursor removal
clear_mz_range = 0.0 0.0 # for iTRAQ/TMT type data; will clear out all peaks in the specified m/z range

#
# additional modifications
#

add_Cterm_peptide = 0.0
add_Nterm_peptide = 0.0
add_Cterm_protein = 0.0
add_Nterm_protein = 0.0

add_G_glycine = 0.0000 # added to G - avg. 57.0513, mono. 57.02146
add_A_alanine = 0.0000 # added to A - avg. 71.0779, mono. 71.03711
add_S_serine = 0.0000 # added to S - avg. 87.0773, mono. 87.03203
add_P_proline = 0.0000 # added to P - avg. 97.1152, mono. 97.05276
add_V_valine = 0.0000 # added to V - avg. 99.1311, mono. 99.06841
add_T_threonine = 0.0000 # added to T - avg. 101.1038, mono. 101.04768
add_C_cysteine = 0.0000 # added to C - avg. 103.1429, mono. 103.00918
add_L_leucine = 0.0000 # added to L - avg. 113.1576, mono. 113.08406
add_I_isoleucine = 0.0000 # added to I - avg. 113.1576, mono. 113.08406
add_N_asparagine = 0.0000 # added to N - avg. 114.1026, mono. 114.04293
add_D_aspartic_acid = 0.0000 # added to D - avg. 115.0874, mono. 115.02694
add_Q_glutamine = 0.0000 # added to Q - avg. 128.1292, mono. 128.05858
add_K_lysine = 0.0000 # added to K - avg. 128.1723, mono. 128.09496
add_E_glutamic_acid = 0.0000 # added to E - avg. 129.1140, mono. 129.04259
add_M_methionine = 0.0000 # added to M - avg. 131.1961, mono. 131.04048
add_H_histidine = 0.0000 # added to H - avg. 137.1393, mono. 137.05891
add_F_phenylalanine = 0.0000 # added to F - avg. 147.1739, mono. 147.06841
add_U_selenocysteine = 0.0000 # added to U - avg. 150.0379, mono. 150.95363
add_R_arginine = 0.0000 # added to R - avg. 156.1857, mono. 156.10111
add_Y_tyrosine = 0.0000 # added to Y - avg. 163.0633, mono. 163.06333

```
add_W_tryptophan = 0.0000            # added to W - avg. 186.0793, mono. 186.07931
add_O_pyrrolysine = 0.0000           # added to O - avg. 237.2982, mono  237.14773
add_B_user_amino_acid = 0.0000       # added to B - avg.   0.0000, mono.   0.00000
add_J_user_amino_acid = 0.0000       # added to J - avg.   0.0000, mono.   0.00000
add_X_user_amino_acid = 0.0000       # added to X - avg.   0.0000, mono.   0.00000
add_Z_user_amino_acid = 0.0000       # added to Z - avg.   0.0000, mono.   0.00000

#
# COMET_ENZYME_INFO _must_ be at the end of this parameters file
#
[COMET_ENZYME_INFO]
0.  Cut_everywhere       0      -           -
1.  Trypsin              1      KR          P
2.  Trypsin/P            1      KR          -
3.  Lys_C                1      K           P
4.  Lys_N                0      K           -
5.  Arg_C                1      R           P
6.  Asp_N                0      D           -
7.  CNBr                 1      M           -
8.  Glu_C                1      DE          P
9.  PepsinA              1      FL          P
10. Chymotrypsin         1      FWYL        P
11. CDAP                 0      C           -
12. No_cut               1      @           @
```

**SI File 2: Comet (TPP6) parameters for N-terminal peptides search**

```
# comet_version 2021.01 rev. 0
# Comet MS/MS search engine parameters file.
# Everything following the '#' symbol is treated as a comment.

database_name = /some/path/db.fasta
decoy_search = 0                    # 0=no (default), 1=concatenated search, 2=separate search
peff_format = 0                     # 0=no (normal fasta, default), 1=PEFF PSI-MOD, 2=PEFF Unimod
peff_obo = C:/TPP/conf/PSI-MOD.obo                          # path to PSI Mod or Unimod OBO file

num_threads = 0                     # 0=poll CPU to set num threads; else specify num threads directly (max 128)

#
# masses
#
peptide_mass_tolerance = 1.1
peptide_mass_units = 0              # 0=amu, 1=mmu, 2=ppm
mass_type_parent = 1                # 0=average masses, 1=monoisotopic masses
mass_type_fragment = 1              # 0=average masses, 1=monoisotopic masses
precursor_tolerance_type = 1        # 0=MH+ (default), 1=precursor m/z; only valid for amu/mmu tolerances
isotope_error = 3                   # 0=off, 1=0/1 (C13 error), 2=0/1/2, 3=0/1/2/3, 4=-8/-4/0/4/8 (for +4/+8 labeling)

#
# search enzyme
#
search_enzyme_number = 12           # choose from list at end of this params file
search_enzyme2_number = 0           # second enzyme; set to 0 if no second enzyme
num_enzyme_termini = 2              # 1 (semi-digested), 2 (fully digested, default), 8 C-term unspecific , 9 N-term
unspecific
allowed_missed_cleavage = 0         # maximum value is 5; for enzyme search

#
# Up to 9 variable modifications are supported
# format:  <mass> <residues> <0=variable/else binary> <max_mods_per_peptide> <term_distance> <n/c-term>
<required> <neutral_loss>
#     e.g. 79.966331 STY 0 3 -1 0 0 97.976896
#
variable_mod01 = 15.9949 M 0 3 -1 0 0 0.0
variable_mod02 = 0.984 DE 0 1 -1 0 0 0.0
max_variable_mods_in_peptide = 3
require_variable_mod = 0

#
# fragment ions
#
# ion trap ms/ms:  1.0005 tolerance, 0.4 offset (mono masses), theoretical_fragment_ions = 1
# high res ms/ms:    0.02 tolerance, 0.0 offset (mono masses), theoretical_fragment_ions = 0,
spectrum_batch_size = 15000
#
```

fragment_bin_tol = 1.0005 # binning to use on fragment ions
fragment_bin_offset = 0.4 # offset position to start the binning (0.0 to 1.0)
theoretical_fragment_ions = 1 # 0=use flanking peaks, 1=M peak only
use_A_ions = 0
use_B_ions = 1
use_C_ions = 0
use_X_ions = 0
use_Y_ions = 1
use_Z_ions = 0
use_Z1_ions = 0
use_NL_ions = 1 # 0=no, 1=yes to consider NH3/H2O neutral loss peaks

#
# output
#
output_sqtfile = 0 # 0=no, 1=yes write sqt file
output_txtfile = 0 # 0=no, 1=yes write tab-delimited txt file
output_pepxmlfile = 1 # 0=no, 1=yes write pepXML file
output_mzidentmlfile = 0 # 0=no, 1=yes write mzIdentML file
output_percolatorfile = 0 # 0=no, 1=yes write Percolator pin file
print_expect_score = 1 # 0=no, 1=yes to replace Sp with expect in out & sqt
num_output_lines = 5 # num peptide results to show

sample_enzyme_number = 11 # Sample enzyme which is possibly different than the one applied to the search.
# Used to calculate NTT & NMC in pepXML output (default=1 for trypsin).

#
# mzXML parameters
#
scan_range = 0 0 # start and end scan range to search; either entry can be set independently
precursor_charge = 0 0 # precursor charge range to analyze; does not override any existing charge; 0 as 1st entry ignores parameter
override_charge = 0 # 0=no, 1=override precursor charge states, 2=ignore precursor charges outside precursor_charge range, 3=see online
ms_level = 2 # MS level to analyze, valid are levels 2 (default) or 3
activation_method = ALL # activation method; used if activation method set; allowed ALL, CID, ECD, ETD, ETD+SA, PQD, HCD, IRMPD, SID

#
# misc parameters
#
digest_mass_range = 500.0 10000.0 # MH+ peptide mass range to analyze
peptide_length_range = 5 63 # minimum and maximum peptide length to analyze (default 1 63; max length 63)
num_results = 100 # number of search hits to store internally
max_duplicate_proteins = -1 # maximum number of additional duplicate protein names to report for each peptide ID; -1 reports all duplicates
max_fragment_charge = 5 # set maximum fragment charge state to analyze (allowed max 5)
max_precursor_charge = 9 # set maximum precursor charge state to analyze (allowed max 9)

```
nucleotide_reading_frame = 0          # 0=proteinDB, 1-6, 7=forward three, 8=reverse three, 9=all six
clip_nterm_methionine = 0             # 0=leave sequences as-is; 1=also consider sequence w/o N-term methionine
spectrum_batch_size = 15000           # max. # of spectra to search at a time; 0 to search the entire scan range in
one loop
decoy_prefix = DECOY_                 # decoy entries are denoted by this string which is pre-pended to each protein
accession
equal_I_and_L = 1                     # 0=treat I and L as different; 1=treat I and L as same
output_suffix =                       # add a suffix to output base names i.e. suffix "-C" generates base-C.pep.xml from
base.mzXML input
mass_offsets =                        # one or more mass offsets to search (values substracted from deconvoluted
precursor mass)
precursor_NL_ions =                   # one or more precursor neutral loss masses, will be added to xcorr analysis

#
# spectral processing
#
minimum_peaks = 10                    # required minimum number of peaks in spectrum to search (default 10)
minimum_intensity = 0                 # minimum intensity value to read in
remove_precursor_peak = 0             # 0=no, 1=yes, 2=all charge reduced precursor peaks (for ETD), 3=phosphate
neutral loss peaks
remove_precursor_tolerance = 1.5      # +- Da tolerance for precursor removal
clear_mz_range = 0.0 0.0              # for iTRAQ/TMT type data; will clear out all peaks in the specified m/z range

#
# additional modifications
#

add_Cterm_peptide = 0.0
add_Nterm_peptide = 0.0
add_Cterm_protein = 0.0
add_Nterm_protein = 0.0

add_G_glycine = 0.0000                # added to G - avg.  57.0513, mono.  57.02146
add_A_alanine = 0.0000                # added to A - avg.  71.0779, mono.  71.03711
add_S_serine = 0.0000                 # added to S - avg.  87.0773, mono.  87.03203
add_P_proline = 0.0000                # added to P - avg.  97.1152, mono.  97.05276
add_V_valine = 0.0000                 # added to V - avg.  99.1311, mono.  99.06841
add_T_threonine = 0.0000              # added to T - avg. 101.1038, mono. 101.04768
add_C_cysteine = 0.0000               # added to C - avg. 103.1429, mono. 103.00918
add_L_leucine = 0.0000                # added to L - avg. 113.1576, mono. 113.08406
add_I_isoleucine = 0.0000             # added to I - avg. 113.1576, mono. 113.08406
add_N_asparagine = 0.0000             # added to N - avg. 114.1026, mono. 114.04293
add_D_aspartic_acid = 0.0000          # added to D - avg. 115.0874, mono. 115.02694
add_Q_glutamine = 0.0000              # added to Q - avg. 128.1292, mono. 128.05858
add_K_lysine = 0.0000                 # added to K - avg. 128.1723, mono. 128.09496
add_E_glutamic_acid = 0.0000          # added to E - avg. 129.1140, mono. 129.04259
add_M_methionine = 0.0000             # added to M - avg. 131.1961, mono. 131.04048
add_H_histidine = 0.0000              # added to H - avg. 137.1393, mono. 137.05891
add_F_phenylalanine = 0.0000          # added to F - avg. 147.1739, mono. 147.06841
add_U_selenocysteine = 0.0000         # added to U - avg. 150.0379, mono. 150.95363
```

```
add_R_arginine = 0.0000              # added to R - avg. 156.1857, mono. 156.10111
add_Y_tyrosine = 0.0000              # added to Y - avg. 163.0633, mono. 163.06333
add_W_tryptophan = 0.0000            # added to W - avg. 186.0793, mono. 186.07931
add_O_pyrrolysine = 0.0000           # added to O - avg. 237.2982, mono  237.14773
add_B_user_amino_acid = 0.0000       # added to B - avg.   0.0000, mono.   0.00000
add_J_user_amino_acid = 0.0000       # added to J - avg.   0.0000, mono.   0.00000
add_X_user_amino_acid = 0.0000       # added to X - avg.   0.0000, mono.   0.00000
add_Z_user_amino_acid = 0.0000       # added to Z - avg.   0.0000, mono.   0.00000

#
# COMET_ENZYME_INFO _must_ be at the end of this parameters file
#
[COMET_ENZYME_INFO]
0.  Cut_everywhere         0      -           -
1.  Trypsin                1      KR          P
2.  Trypsin/P              1      KR          -
3.  Lys_C                  1      K           P
4.  Lys_N                  0      K           -
5.  Arg_C                  1      R           P
6.  Asp_N                  0      D           -
7.  CNBr                   1      M           -
8.  Glu_C                  1      DE          P
9.  PepsinA                1      FL          P
10. Chymotrypsin           1      FWYL        P
11. CDAP                   0      C           -
12. No_cut                 1      @           @
```